\documentclass{emulateapj}

\usepackage{amsmath}
\usepackage{color}
\usepackage{graphicx}
\usepackage{natbib}
\usepackage{booktabs}

\newcommand{\pz}{\phantom{0}}
\newcommand{\pp}{\phantom{$+$}}

\shorttitle{RadioAstron polarimetric space VLBI observations of BL~Lacertae}
\shortauthors{G\'omez et al.}

\begin{document}

\title{Probing the innermost regions of AGN jets and their magnetic fields with {\it RadioAstron}. I. Imaging BL~Lacertae at 21 microarcsecond resolution}

%% Authors
\author{Jos\'e L. G\'omez\altaffilmark{1}, Andrei P. Lobanov\altaffilmark{2,3}, Gabriele Bruni\altaffilmark{2}, Yuri Y. Kovalev\altaffilmark{4,2}, Alan P. Marscher\altaffilmark{5}, Svetlana G. Jorstad\altaffilmark{5,6}, Yosuke Mizuno\altaffilmark{7}, Uwe Bach\altaffilmark{2}, Kirill V. Sokolovsky\altaffilmark{4,8,9}, James M. Anderson\altaffilmark{2,10},
Pablo Galindo\altaffilmark{1}, Nikolay S. Kardashev\altaffilmark{4}, and Mikhail M. Lisakov\altaffilmark{4}}

\altaffiltext{1}{Instituto de Astrof\'{\i}sica de Andaluc\'{\i}a-CSIC, Glorieta de la Astronom\'{\i}a s/n, 18008 Granada, Spain. jlgomez@iaa.csic.es}

\altaffiltext{2}{Max-Planck-Institut f\"ur Radioastronomie, Auf dem H\"ugel 69, 53121 Bonn, Germany}

\altaffiltext{3}{Institut f\"ur Experimentalphysik, Universit\"at Hamburg, Luruper Chaussee 149, 22761 Hamburg, Germany}

\altaffiltext{4}{Astro Space Center, Lebedev Physical Institute, Russian Academy of Sciences, Profsoyuznaya str. 84/32, Moscow 117997, Russia}

\altaffiltext{5}{Institute for Astrophysical Research, Boston University, 725 Commonwealth Avenue, Boston, MA 02215}

\altaffiltext{6}{Astronomical Institute, St.~Petersburg State University, Universitetskij Pr.~28, Petrodvorets, 198504 St.Petersburg, Russia}

\altaffiltext{7}{Institute for Theoretical Physics, Goethe University, 60438, Frankfurt am Main, Germany}

\altaffiltext{8}{Sternberg Astronomical Institute, Moscow State University, Universitetskii pr. 13, 119992 Moscow, Russia}

\altaffiltext{9}{Institute of Astronomy, Astrophysics, Space Applications and Remote Sensing, National Observatory of Athens, Vas.~Pavlou~\&~I.~Metaxa, GR-15~236 Penteli, Greece}

\altaffiltext{10}{Helmholtz-Zentrum Potsdam, Deutsches GeoForschungsZentrum GFZ, Telegrafenberg, 14473 Potsdam, Germany}

\begin{abstract}
\noindent
We present the first polarimetric space VLBI imaging observations at 22~GHz. BL~Lacertae was observed in 2013 November 10 with the \emph{RadioAstron} space VLBI mission, including a ground array of 15 radio telescopes. The instrumental polarization of the space radio telescope is found to be within 9\%, demonstrating the polarimetric imaging capabilities of \emph{RadioAstron} at 22~GHz. Ground--space fringes were obtained up to a projected baseline distance of 7.9~Earth's diameters in length, allowing us to image the jet in BL~Lacertae with a maximum angular resolution of 21~$\mu$as, the highest achieved to date. We find evidence for emission upstream of the radio core, which may correspond to a recollimation shock at about 40~$\mu$as from the jet apex, in a pattern that includes other recollimation shocks at approximately 100~$\mu$as and 250~$\mu$as from the jet apex. Polarized emission is detected in two components within the innermost 0.5~mas from the core, as well as in some knots 3 mas downstream. Faraday rotation analysis, obtained from combining \emph{RadioAstron} 22~GHz and ground-based 15~GHz and 43~GHz images, shows a gradient in rotation measure and Faraday corrected polarization vector as a function of position angle with respect to the core, suggesting that the jet in BL~Lacertae is threaded by a helical magnetic field. The intrinsic de-boosted brightness temperature in the unresolved core exceeds $3\!\times\!10^{12}$~K, suggesting at the very least departure from equipartition of energy between the magnetic field and radiating particles.
\end{abstract}


\section{Introduction}

  Accretion of gas onto the supermassive black holes lurking at the center of active galactic nuclei (AGN) gives rise to powerful relativistic jets \citep[e.g.,][]{2002Natur.417..625M}. These are produced by dynamically important magnetic fields twisted by differential rotation of the black hole's accretion disk or ergosphere \citep{1977MNRAS.179..433B,Blandford:1982vy,McKinney:2009ko,Zamaninasab:2014hs}. Observational signatures for the existence of such helical magnetic fields can be obtained by looking for Faraday rotation gradients, produced by the systematic change in the net line-of-sight magnetic field component across the jet width \citep{Laing:1981bx,Asada:2002to}.
  
  Obtaining a better understanding of the jet formation, and of the role played by the magnetic field requires probing the innermost regions of AGN jets, but this is limited by the insufficient angular resolution provided by existing, ground-based, very long baseline interferometry (VLBI) arrays. However, space VLBI, in which one of the antennas is in Earth orbit, is capable of extending the baseline distances beyond the Earth's diameter, reaching unprecedentedly high angular resolutions in astronomical observations \citep[e.g.,][]{1986Sci...234..187L,2000ApJ...530..245G,2001MNRAS.320L..49G,Lobanov:2001ju}.
  
  On 2011 July 18, the \emph{RadioAstron} space VLBI mission \citep{2013ARep...57..153K} began to operate, featuring a 10 m space radio telescope (SRT) on board the satellite \emph{Spektr-R}. \emph{RadioAstron} provides the first true full-polarization capabilities for space VLBI observations on baselines longer than the Earth's diameter at 0.32, 1.6, and 22~GHz. The SRT operates also at 5~GHz, but an onboard hardware failure limits the recording mode at this frequency to left circular polarization only.

   The current paradigm for AGN is that their radio emission is explained by synchrotron radiation from relativistic electrons that are Doppler boosted through bulk motion. In this model, the intrinsic brightness temperatures cannot exceed $10^{11}$ to $10^{12}$~K \citep{1969ApJ...155L..71K,1994ApJ...426...51R}. Typical Doppler boosting is expected to be able to raise this temperature by a factor of $\sim10$ \citep[see also][]{2009A&A...494..527H,Lister:2013gp}. For direct interferometric measurements, increasing the interferometer baseline length is the only way to measure higher brightness temperatures \citep[see e.g.,][]{2005AJ....130.2473K}, and hence, to place stringent observational constraints on the physics of the most energetic relativistic outflows. The highest observing frequency, 22~GHz, and resolution of \emph{RadioAstron} allow us to probe the most energetic regions located closer to the central engine \cite[see, e.g.,][]{Lobanov:1998vr,Sokolovsky:2011bl,Pushkarev:2012cr} while scattering effects in the Galaxy are negligible \citep{2015MNRAS.452.4274P}.
  
  First polarimetric space VLBI imaging observations with \emph{RadioAstron} were performed on 2013 March 9 during the early science program, targeting the high-redshift quasar TXS~0642+449 at a frequency of 1.6~GHz \citep{2015A&A...583A.100L}. Instrumental polarization of the SRT was found to be smaller than 9\% in amplitude, demonstrating the polarimetric imaging capabilities of \emph{RadioAstron} at this frequency \citep[see also][]{Pashchenko:2015gh}. Fringes on ground-space baselines were found up to projected baseline distances of 6 Earth's diameters in length, allowing imaging of 0642+449 with an angular resolution of 0.8 mas -- a four-fold improvement over ground VLBI observations at this frequency.
  
  In this paper we present the first polarimetric space VLBI imaging observations at 22~GHz, obtained as part of our \emph{RadioAstron} Key Science Program (KSP), aimed to develop, commission, and exploit the unprecedented high angular resolution polarization capabilities of \emph{RadioAstron} to probe the innermost regions of AGN jets and their magnetic fields. A sample of powerful, highly polarized, and $\gamma$-ray emitting blazars is being observed within our KSP, including several quasars, BL~Lac objects, and radio galaxies. In this first paper of a series containing our \emph{RadioAstron} KSP results, we present our observations of BL~Lacertae, the eponymous blazar that gives name to the class of BL~Lac objects.
  
  The jet of BL~Lac is pointing at us with a viewing angle of $\sim$8$^{\circ}$ with bulk flow at a Lorentz factor of $\sim$7 \citep{2005AJ....130.1418J}. Previous observations have revealed a multiwavelength outburst, from radio to $\gamma$-rays, triggered by the passing of a bright moving feature through a standing shock associated with the core of the jet \citep{Marscher:2008ii}. Rotation of the optical polarization angle prior to the $\gamma$-ray flare led \cite{Marscher:2008ii} to conclude that the acceleration and collimation zone, upstream of the radio core, is threaded by a helical magnetic field. The existence of a helical magnetic field in BL~Lac has also been suggested by \cite{2015ApJ...803....3C} through the analysis of the MOJAVE monitoring program data. These authors claim that Alfv\'en waves triggering the formation of superluminal components are excited by changes in the position angle of a recollimation shock (located at a distance from the core of $\sim$0.26 mas), in a similar way as found in magnetohydrodynamical simulations of relativistic jets threaded by a helical magnetic field \citep{1989ApJ...344...89L,2013EPJWC..6101001M}.
  
  VLBA monitoring programs at 43~GHz have revealed the existence of a second stationary feature besides the one described previously at 0.26 mas, located at a distance of $\sim$0.1 mas \citep{2005AJ....130.1418J}. Variations in the position angle of these innermost components suggest that the jet in BL~Lac may be precessing \citep{2003MNRAS.341..405S,2005ApJ...623...79M}.
  
  Faraday rotation analysis reveals a variable rotation measure (RM) in the core of BL~Lac, including sign reversals. Observations by \cite{Zavala:2003eq} give $-376$ rad m$^{-2}$, while \cite{OSullivan:2009bx} find RM values between $-1000$ and $+240$ rad m$^{-2}$, depending also on the set of frequencies used in the analysis. Significantly larger positive values, between approximately 2000 and 10000 rad m$^{-2}$, are reported by \cite{2003MNRAS.341..405S} and \cite{2007AJ....134..799J}. However rotation measure values for the jet appear very stable, with smaller values in the range between $-300$ and $\sim$150 rad m$^{-2}$ \citep{Zavala:2003eq,OSullivan:2009bx,Hovatta:2012jv}.

  In Sec.~\ref{Sec:Obs} we present the observations and the specific details for the analysis of the space VLBI \emph{RadioAstron} data; in Sec.~\ref{Sec:im} we present and analyze the \emph{RadioAstron} image at 22~GHz, whose polarization is analyzed in more detail in Sec.~\ref{Sec:pol}. Finally, our conclusions and summary are presented in Sec.~\ref{Sec:sum}.
  
  For a flat Universe with $\Omega_{m}$= 0.3, $\Omega_{\Lambda}$= 0.7, and $H_{0}$ = 70 km s$^{-1}$ Mpc$^{-1}$ \citep{2014A&A...571A..16P}, 1 mas corresponds to {1.295} pc at the redshift of BL~Lac ($z$=0.0686), and a proper motion of 1 mas/yr is equivalent to $4.51\,c$.

\section{Observations and data reduction}
\label{Sec:Obs}

\subsection{{\it RadioAstron} space VLBI observations at 22~GHz}
\label{Sec:RA_data}
  \emph{RadioAstron} observations of BL~Lac at 22.2~GHz (K-band) were performed on 2013 November 10--11 (from 21:30 to 13:00 UT). A total of 26 ground antennas were initially scheduled for the observations, but different technical problems at the sites limited the final number of correlated ground antennas to 15, namely Effelsberg (EF), Metsaehovi (MH), Onsala (ON), Svetloe (SV), Zelenchukskaya (ZC), Medicina (MC), Badary (BD), and VLBA antennas Brewster (BR), Hancock (HN), Kitt Peak (KP), Los Alamos (LA), North Liberty (NL), Owens Valley (OV), Pie Town (PT), and Mauna Kea (MK).

  The data were recorded in two polarizations (left and right circularly polarized, LCP and RCP), with a total bandwidth of 32 MHz per polarization, split into two intermediate frequency (IF) bands of 16 MHz. The SRT data were recorded by the \emph{RadioAstron} satellite tracking stations \citep{2013ARep...57..153K,2014SPIE.9145E..0BF} in Puschino (21:30\,--\,06:10 UT) and Green Bank (07:30\,--\,13:00 UT), including some extended gaps required for cooling of the motor drive of the onboard high-gain antenna of the \emph{Spektr-R} satellite. These gaps were used for the ground-only observations of BL~Lac at 15~GHz and 43~GHz (see Sec.~\ref{Sec:U-Q}), as well as observations of several calibrator sources.

  Correlation of the data was performed using the upgraded version of the DiFX correlator developed at the Max-Planck-Institut f\"ur Radiostronomie (MPIfR) in Bonn \citep{2014evn..confE.119B}, enabling accurate calibration of the instrumental polarization of the SRT \citep[see also][for a more detailed description of the imaging and correlation of \emph{RadioAstron} observations]{2015A&A...583A.100L}.
  
  The correlated data were reduced and imaged using a combination of the AIPS and \emph{Difmap} \citep{Shepherd:1997wv} software packages. The a priori amplitude calibration was applied using the measured system temperatures for the ground antennas and the SRT. Sensitivity parameters of the SRT \citep{2014CosRe..52..393K} are measured regularly during maintenance sessions. Parallactic angle corrections were applied to the ground antennas to correct for the feed rotation.

\begin{figure}[t]
\epsscale{1.15}
\plotone{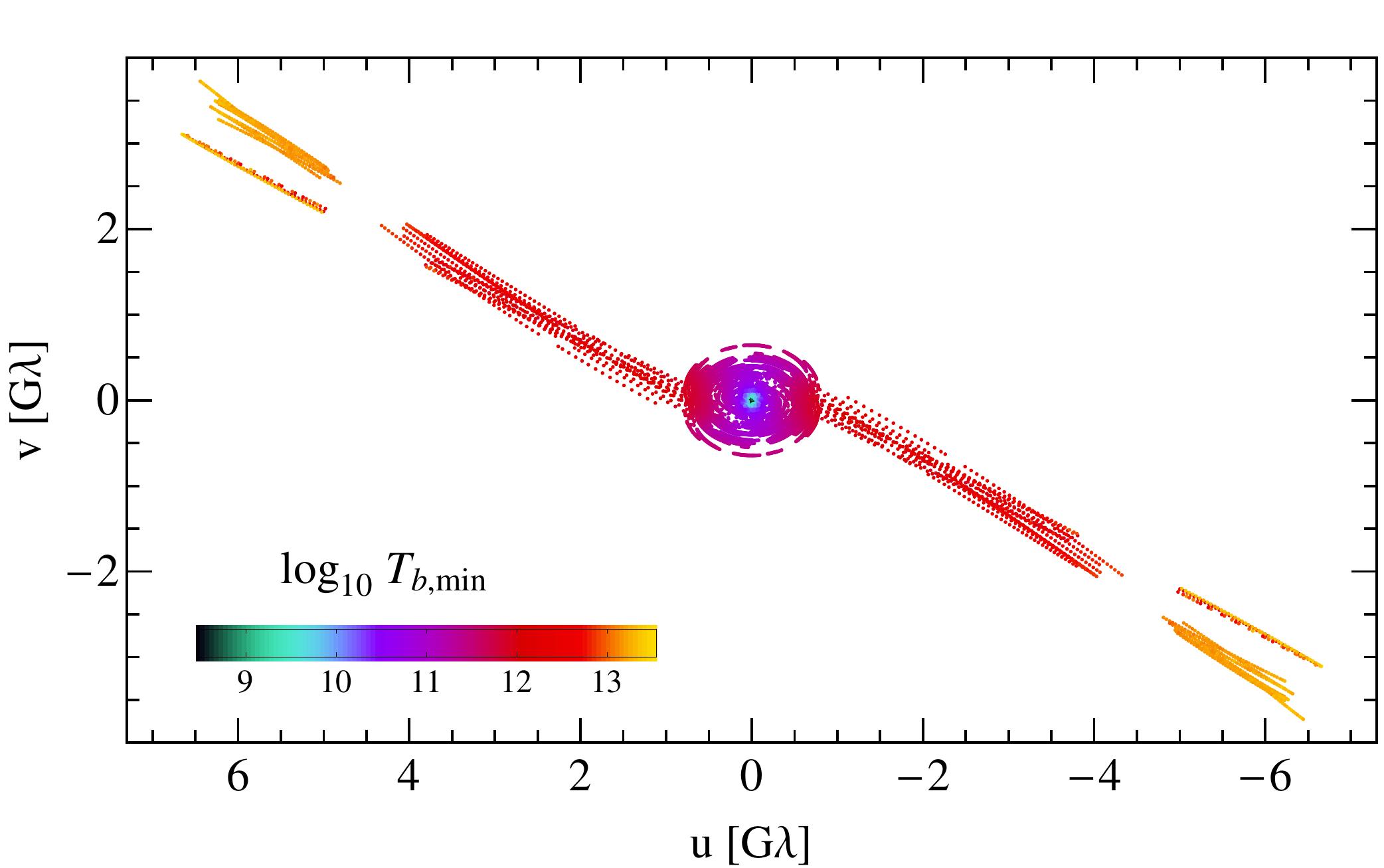}
\caption{Fourier coverage (\emph{uv}-coverage) of the fringe fitted data (i.e., reliable fringe detections) of the \emph{RadioAstron} observations of BL~Lac on 2013 November 10-11 at 22~GHz. Color marks the lower limit of observed brightness temperature obtained from visibility amplitudes (for details, see discussion in Sect.~\ref{Sec:Tb}).}
\label{Fig:uvplot}
\end{figure}

\subsubsection{Fringe fitting}
  Fringe fitting of the data was performed by first manually solving for the instrumental phase offsets and single band delays using a short scan during the perigee of the SRT, when the shortest projected ground--space baseline distances (smaller than one Earth's diameter) are obtained, providing the best fringe solutions for the space antenna. These solutions were applied before performing a global fringe search for the delays and rates of the ground array only.

  Once the ground antennas were fully calibrated, they were coherently combined to improve the fringe detection sensitivity of the SRT \citep{Kogan:1996va}. This baseline stacking was carried out setting DPARM(1)=3 in AIPS's FRING task, performing also an exhaustive baseline search, and combining both polarizations and IFs to improve the sensitivity. Progressively longer solution intervals were used for the fringe search, from one minute to maintain coherence of the signal during the acceleration of the space craft in the perigee, to four minutes to increase the sensitivity on the longer baselines to the SRT. 

\begin{table}[t!]
\caption{Instrumental polarization (D-terms) at 22~GHz}
\label{Tb:dterms}
\begin{center}
\begin{tabular}{c|rr|rr}\hline\hline
Antenna & \multicolumn{2}{|c|}{RCP} &\multicolumn{2}{c}{LCP} \\ \cline{2-5} 
        & $m$  & $\chi$     & $m$  & $\chi$     \\
        & [\%] & [$^\circ$] & [\%] & [$^\circ$] \\ \hline
SRT     &  9.3$\pm$0.5  &   $21\pm$5\pz   &  4.5$\pm$0.3  &   $72\pm$5\pz \\
        &  8.8$\pm$0.8  &   $20\pm$4\pz   &  4.4$\pm$0.2  &   $68\pm$8\pz \\
        \hline
BR      &  1.4$\pm$0.7  &  $-73\pm$18     &  0.8$\pm$0.4  & $-165\pm$22   \\
        &  1.4$\pm$0.7  &  $-86\pm$23     &  0.7$\pm$0.3  & $-196\pm$24   \\
        \hline
EF      &  9.9$\pm$0.7  &  $-91\pm$4\pz   &  8.1$\pm$0.5  & $-126\pm$7\pz \\
        &  9.3$\pm$0.8  &  $-98\pm$3\pz   &  7.5$\pm$0.3  & $-130\pm$6\pz \\
        \hline
HN      &  2.3$\pm$0.2  &  $174\pm$16     &  2.2$\pm$0.6  &   $90\pm$6\pz \\
        &  2.2$\pm$0.4  &  $149\pm$14     &  2.0$\pm$0.8  &   $85\pm$11   \\
        \hline
KP      &  1.1$\pm$0.3  & $-160\pm$8\pz   &  1.1$\pm$0.4  & $-167\pm$12   \\
        &  0.9$\pm$0.4  & $-182\pm$5\pz   &  1.0$\pm$0.3  & $-194\pm$8\pz \\
        \hline
LA      &  2.4$\pm$0.6  &  $-62\pm$7\pz   &  1.0$\pm$0.4  & $-124\pm$8\pz \\
        &  2.7$\pm$0.7  &  $-75\pm$6\pz   &  0.8$\pm$0.6  & $-126\pm$11   \\
        \hline
NL      &  3.6$\pm$0.5  &  $-43\pm$5\pz   &  3.8$\pm$0.2  & $-106\pm$9\pz \\
        &  3.1$\pm$0.6  &  $-43\pm$7\pz   &  4.0$\pm$0.4  & $-104\pm$8\pz \\
        \hline
OV      &  2.0$\pm$0.8  &  $118\pm$8\pz   &  2.3$\pm$0.4  &   $18\pm$7\pz \\
        &  2.2$\pm$0.8  &   $94\pm$6\pz   &  2.7$\pm$0.4  &   $13\pm$11   \\
        \hline
PT      &  1.4$\pm$0.3  &  $-85\pm$11     &  2.0$\pm$0.5  &  $-74\pm$9\pz \\
        &  1.4$\pm$0.4  &  $-77\pm$12     &  2.0$\pm$0.6  &  $-65\pm$12   \\
        \hline
MH      &  1.9$\pm$0.9  & $-116\pm$8\pz   &  8.2$\pm$1.3  &  $-41\pm$9\pz \\
        &  2.9$\pm$0.8  & $-152\pm$6\pz   &  5.5$\pm$0.9  &  $-57\pm$8\pz \\
        \hline
ON      &  4.1$\pm$0.8  &  $-43\pm$4\pz   &  5.3$\pm$0.5  &  $-87\pm$7\pz \\
        &  4.2$\pm$0.9  &  $-38\pm$8\pz   &  5.2$\pm$0.6  &  $-82\pm$12   \\
        \hline
SV      &  4.8$\pm$0.3  &  $179\pm$7\pz   &  4.1$\pm$0.5  &   $-2\pm$5\pz \\
        &  4.7$\pm$0.2  &  $178\pm$10     &  4.0$\pm$0.3  &    $0\pm$4\pz \\
        \hline
ZC      &  6.1$\pm$0.7  &  $162\pm$13     &  8.3$\pm$0.8  &  $168\pm$9\pz \\
        &  8.3$\pm$0.6  &  $111\pm$10     &  6.9$\pm$0.9  &  $165\pm$5\pz \\
        \hline
MC      &  0.9$\pm$0.7  &  $109\pm$10     &  6.1$\pm$0.6  &   $41\pm$8\pz \\
        &  2.6$\pm$0.9  &   $87\pm$11     &  5.5$\pm$0.9  &   $42\pm$11   \\
        \hline
MK      &  3.3$\pm$0.4  &  $-64\pm$8\pz   &  2.9$\pm$0.3  & $-135\pm$5\pz \\
        &  3.5$\pm$0.6  &  $-67\pm$12     &  2.9$\pm$0.5  & $-130\pm$9\pz \\
        \hline
BD      &  7.7$\pm$0.8  &  $-99\pm$7\pz   &  6.4$\pm$0.6  & $-179\pm$6\pz \\
        &  8.1$\pm$0.9  & $-103\pm$8\pz   &  6.7$\pm$0.8  & $-178\pm$9\pz \\
        \hline
\end{tabular}
\end{center} {\bf Notes.} Listed values correspond to the fractional amplitude, $m$, and phase, $\chi$, of the instrumental polarization for each antenna and polarization in the first (top row) and second (bottom row) IF.
\end{table}

  The phasing of a group of ground-based antennas allowed us to obtain reliable ground--space fringe detections up to projected baselines of 7.9 Earth's diameters ($D_\text{E}$) in length, covering the duration of the experiment within which Puschino was used as the tracking station. No further fringes were obtained to the space craft once the tracking station changed to Green Bank, which is presumably due to a difference in clock setting between the two tracking stations. These were searched for by introducing trial clock offsets for the Green Bank tracking station, and performing new test correlations with a larger fringe-search window of up to 1024 channels and 0.1 sec of integration time in width. However, no further fringes were detected to the SRT. We also note that 1.5 hours passed between the last Puschino scan and the first Green Bank scan, thus increasing the space baseline length and perhaps reducing the correlated flux density below the sensitivity threshold. The obtained fringe-fitted data visibility coverage of the Fourier domain (\emph{uv}-coverage) is shown in Fig.~\ref{Fig:uvplot}.

   After the fringe fitting the delay difference between the two polarizations was corrected using AIPS's task RLDLY, and a complex bandpass function was solved for the receiver.

\subsubsection{Polarization calibration}
  The instrumental polarization (D-terms) was obtained using AIPS's task LPCAL \citep{Leppanen:1995eg}. Table \ref{Tb:dterms} lists the instrumental polarization derived for each telescope, with Effelsberg as the reference antenna. Errors in the instrumental polarization were estimated from the dispersion in the values obtained by performing independent data reductions while using different reference antennas, as well as comparison with values obtained for calibrator sources (2021+614 and 1823+568). Estimated values for the ground antennas are also subject to antenna performance and weather conditions at the sites, which in some cases may lead to larger than usual instrumental polarization values. Amplitude and phase stability across the two IFs confirms the reliability of the estimated values.
  
  Instrumental polarization of the SRT at 22~GHz is within 9\% (5\% for LCP), and remarkably consistent across the two IFs, demonstrating its robust polarization capabilities for \emph{RadioAstron} imaging observations at its highest observing frequency of 22~GHz.
  
  Absolute calibration of the electric vector position angle (EVPA) was obtained from comparison with simultaneous single dish observations at the Effelsberg telescope of our target and calibrator sources. We estimate the error in the EVPA calibration to lie between 5$^{\circ}$ and 10$^{\circ}$.

\subsection{Ground-array observations at 15~GHz and 43~GHz}
\label{Sec:U-Q}

\begin{figure}[t]
\epsscale{1.15}
\plotone{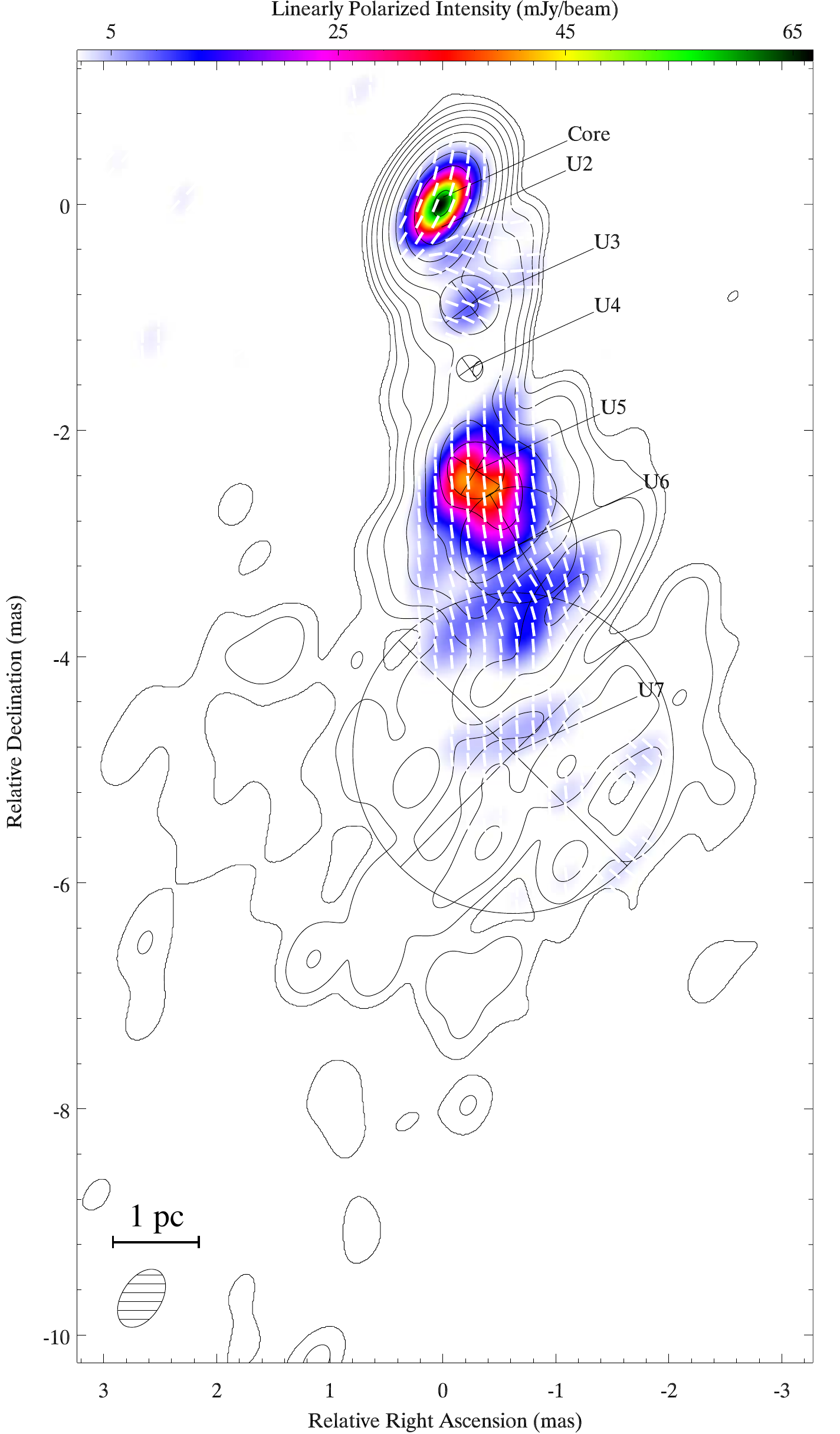}
\caption{Ground-array 15~GHz observations of BL~Lac on 2013 November 10. Total intensity contours are plotted at 0.05, 0.11, 0.24, 0.57, 1.32, 3.08, 7.16, 16.64, 38.70, and 90\% of the peak intensity at 3.31 Jy/beam. Linearly polarized intensity is shown in colors starting at 2 mJy/beam, and bars indicate the EVPA. Synthesized beam FWHM is 0.57$\times$0.35 mas at a position angle of $-33^{\circ}$. Model fit components are also shown overlaid, indicating their position and angular size (see also Table \ref{Tb:modfit}).}
\label{Fig:U-band}
\end{figure}

\begin{figure}[t]
\epsscale{1.15}
\plotone{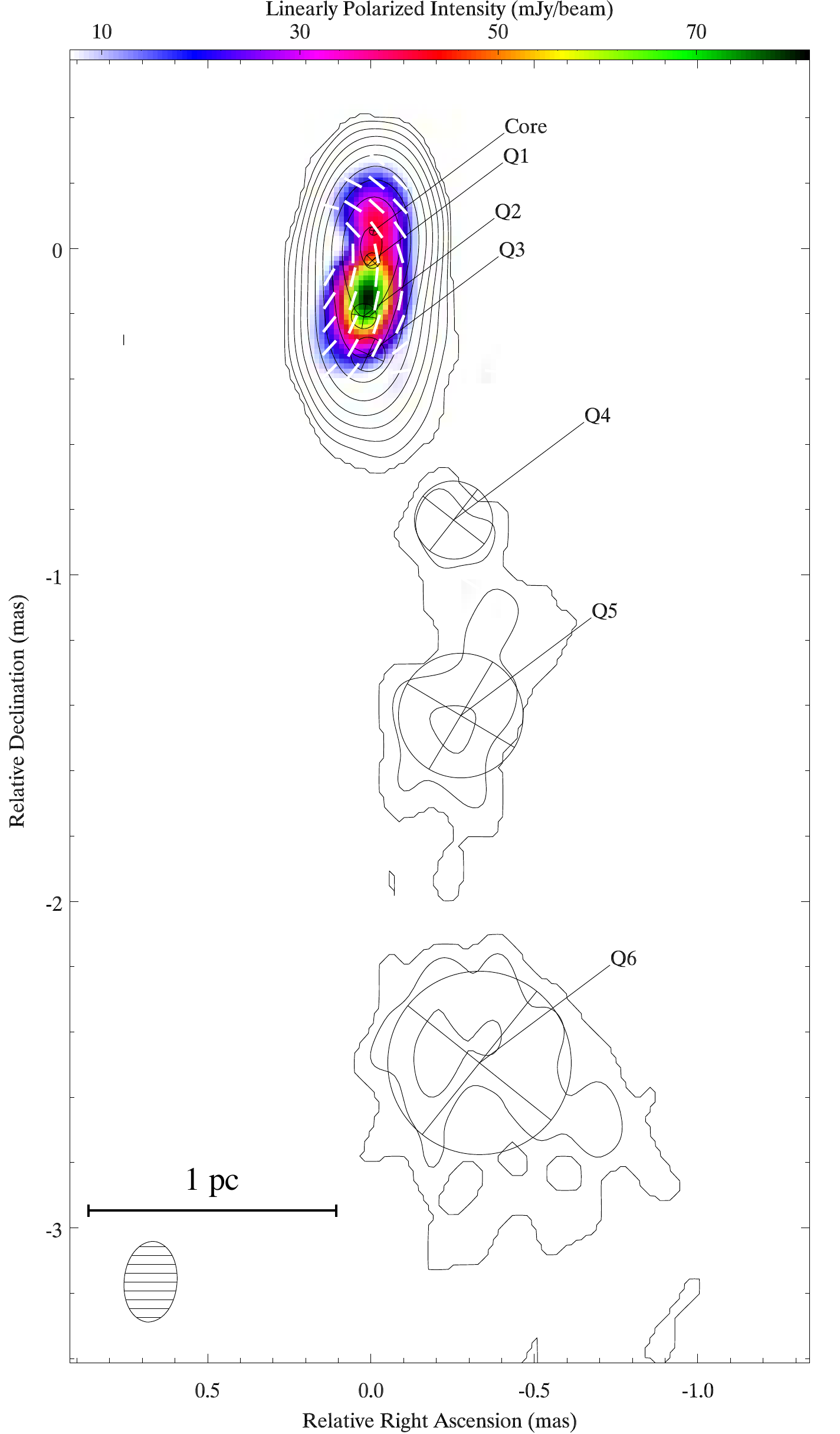}
\caption{VLBA-BU-BLAZAR image of BL~Lac at 43~GHz obtained in 2013 November 18. Total intensity contours are plotted at 0.25, 0.48, 0.92, 1.78, 3.42, 6.57, 12.65, 24.33, 46.79, and 90\% of the peak at 2.7 Jy/beam. Linearly polarized intensity is shown in colors starting at 6.8 mJy/beam, and bars indicate the EVPA. Synthesized beam FWHM is 0.25$\times$0.16 mas at a position angle of $-5^{\circ}$. Model fit components are also shown overlaid, indicating their position and angular size (see also Table \ref{Tb:modfit}).}
\label{Fig:Q-band}
\end{figure}

  Simultaneous ground-only observations of BL~Lac at 15.4~GHz and 43.1~GHz were obtained during gaps in the \emph{RadioAstron} observations. Participating antennas were Effelsberg and VLBA antennas BR, HN, KP, LA, NL, OV, and PT. However, at 43~GHz no fringes were detected on the intercontinental baselines with Effelsberg due to technical problems, severely limiting the sensitivity and angular resolution, in fact preventing the detection of polarization at this frequency. For this reason we have used instead the 43~GHz data from the VLBA-BU-BLAZAR\footnote{see \url{https://www.bu.edu/blazars/VLBAproject.html}} monitoring program, performed on 2013 November 18th, only one week after our observations.
 
\begin{figure*}
\epsscale{0.95}
\plotone{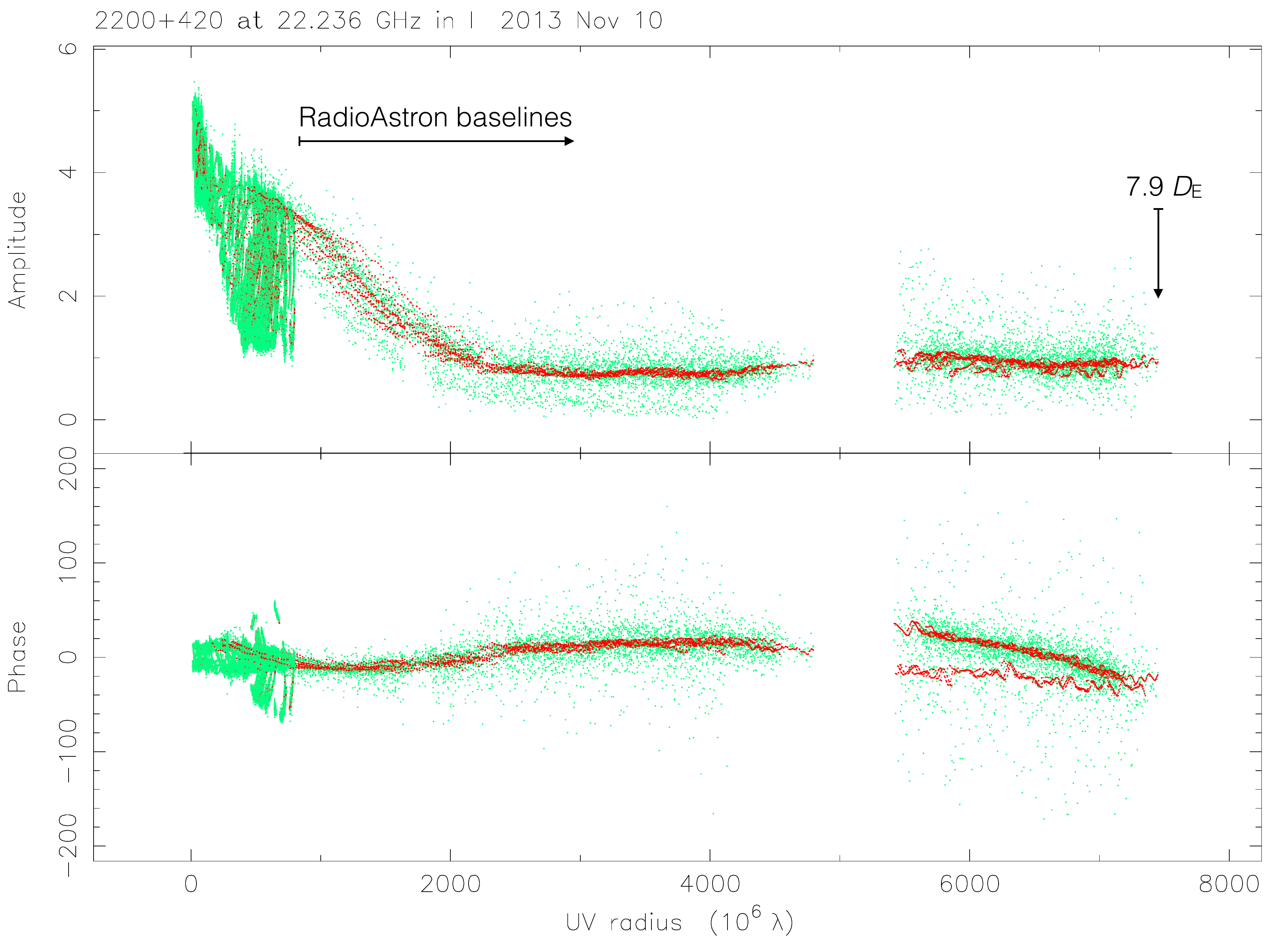}
\caption{Self-calibrated visibility amplitudes and phases as a function of \emph{uv}-distance of the \emph{RadioAstron} observations of BL~Lac on 2013 November 10--11 at 22~GHz. Overplotted in red is the fit to these data of the CLEAN model obtained from the hybrid mapping. Reliable space--ground fringe detections were obtained up to a projected baseline distance of 7.9 Earth's diameters.}
\label{Fig:radplot}
\end{figure*}
  
  Calibration of the 15~GHz data follows that outlined previously in Sec.~\ref{Sec:RA_data}, except for the particular steps related more directly to the space VLBI observations, such as the phasing of the ground-array during the fringe fitting. Calibration of the absolute orientation of the EVPAs was also performed by comparison with Effelsberg single dish observations of BL~Lac ($S_\mathrm{15GHz}=5.57\pm0.23$ Jy, $m=3.44\pm0.3$\%, EVPA=$3.7\pm1.1^{\circ}$), with an estimated error of 5$^{\circ}$ to 7$^{\circ}$.

  Figures \ref{Fig:U-band} and \ref{Fig:Q-band} show the 15~GHz and 43~GHz images of BL~Lac obtained with the ground array. To characterize the emission structure we have performed a fit of the complex visibilities by a set of components with circular Gaussian brightness distributions, listed in Table~\ref{Tb:modfit}.
   
\section{Space VLBI polarimetric images of BL~Lac at 21~$\mu$as angular resolution}
\label{Sec:im} 
  
  Fully calibrated \emph{RadioAstron} data were exported to \emph{Difmap} and imaged using the standard hybrid imaging and self-calibration techniques. Self-calibrated Stokes I visibility amplitudes and phases as a function of Fourier spacing (\emph{uv} distance) and CLEAN model fit to these data are shown in Fig.~\ref{Fig:radplot}. Space VLBI fringes to the SRT extend the projected baseline spacing up to 7.9~$D_\text{E}$, increasing accordingly the angular resolution with respect to that provided by ground-based arrays. However the large eccentricity of the SRT orbit (see Fig.~\ref{Fig:uvplot}) leads to a highly elliptical observing beam.

\begin{figure*}[t]
\epsscale{1.18}
\plotone{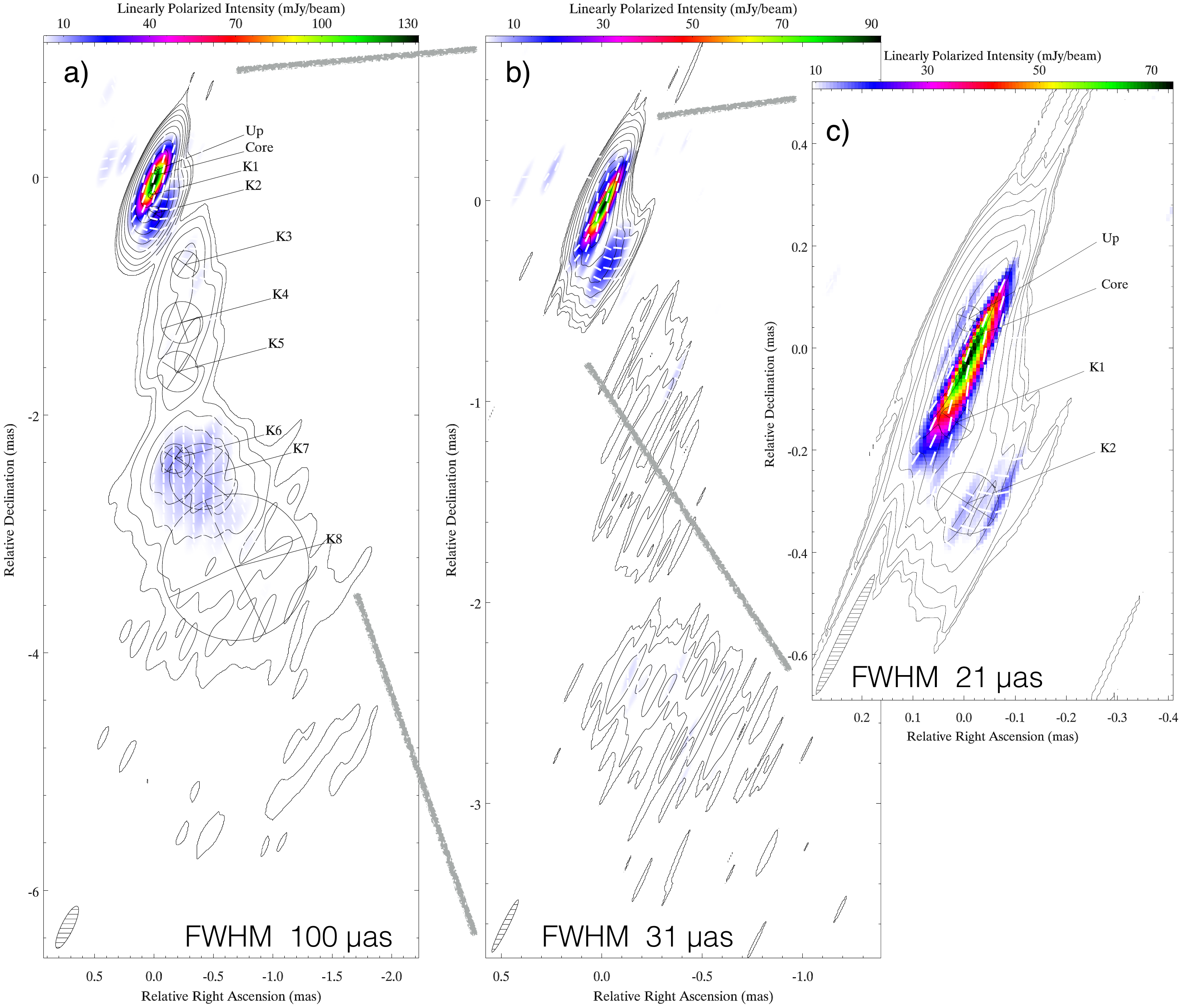}
\caption{\emph{RadioAstron} polarimetric space VLBI images of BL~Lac obtained in 2013 November 10--11 at 22~GHz. From left to right, images are obtained with natural (a), uniform (b), and ``super''-uniform (c) weightings. Total intensity contours are plotted at 0.08, 0.17, 0.36, 0.80, 1.76, 3.86, 8.49, 18.65, 40.97, and 90\% (0.2, 0,4, 0.78, 1.54, 3.04, 5.99, 11.79, 23.22, 45,71, and 90\%; 0.76, 1.3, 2.21, 3.75, 6.37, 10.81, 18.37, 31.2, 23, and 90\%) of the peak intensity at 2.48 (1.48; 1.23) Jy/beam for the natural weighted image (uniform; ``super''-uniform). Linearly polarized intensity is shown in colors starting at (a) 3.1 mJy/beam, (b) 3.8 mJy/beam, and (c) 9.4 mJy/beam, and white bars indicate the (\emph{uncorrected} for Faraday rotation) EVPAs. Synthesized beams are (a) 0.392$\times$0.100~mas, (b) 0.292$\times$0.031~mas, and (c) 0.261$\times$0.021~mas at a common position angle of $-26^{\circ}$.}
\label{Fig:RA_BLLac}
\end{figure*}
  
  \emph{RadioAstron} space VLBI polarimetric images of BL~Lac are shown in Fig.~\ref{Fig:RA_BLLac} for three different visibility weights: natural, uniform, and ``super'' uniform (in which the gridding weights are not scaled by the visibility amplitude errors). The weights of the longest space VLBI visibilities are therefore increasingly higher from natural to super uniform weightings, consequently yielding higher angular resolutions, albeit with lower image sensitivities. The super uniform weighting image yields an angular resolution of 21~$\mu$as (along the minor axis of the restoring beam), which to our knowledge corresponds to the {\it highest achieved to date}. For an estimated black hole mass of $\sim1.6\!\times\!10^8$~M$_{\odot}$ \citep{Woo:2002kw}, where M$_{\odot}$ is the mass of the Sun, this corresponds to a linear resolution of $\sim$1800 Schwarzschild radii at the BL~Lac distance.

  The images in Fig.~\ref{Fig:RA_BLLac} show the familiar radio continuum emission structure of BL~Lac, dominated by the core and a jet that extends to the south. Conveniently, the highest angular resolution provided by the ground-space baselines is obtained nearly along the jet direction, allowing close examination of the innermost structure of the jet. As can be better distinguished in the uniform (Fig.~\ref{Fig:RA_BLLac}b) and super uniform (Fig.~\ref{Fig:RA_BLLac}c) images, the total intensity images reveal a bent structure in the innermost 0.5~mas region. The linearly polarized images clearly distinguish two components in this region, as well as in the jet area at $\sim$3 mas from the core.

\begin{table*}[t!]
\caption{Gaussian model fits for the 22~GHz RadioAstron and ground-array data at 43 and 15~GHz}
\label{Tb:modfit}
\begin{center}
\begin{tabular}{c|c|c|c|c|c|c|c}\hline
Comp. &  Flux       & Distance        & PA           & Size            & T$_\text{b}$  & $m$      & EVPA\\
      & (mJy)       & (mas)           & ($^{\circ}$) & (mas)           & (K)    & (\%)     & ($^{\circ}$)\\
   \hline
   \hline
\multicolumn{8}{c}{RadioAstron 22~GHz} \\
   \hline
Up    &      1578$\pm$72\pz  & 0.041$\pm$0.003     & \pz\pz$-1\pm$4   & 0.050$\pm$0.003     & (1.56$\pm$0.26$)\times10^{12}$    & \pz 4.0$\pm$0.4 &    $-17\pm$1  \\
Core  & \pz   802$\pm$43\pz  & \ldots              & \ldots           & $<$0.01             & $>$2.0$\times10^{13}$             & \pz 4.8$\pm$0.4 &    $-16\pm$1  \\
K1    &      1128$\pm$37\pz  & 0.164$\pm$0.004     & \pp  $172\pm$1   & 0.067$\pm$0.003     & (5.82$\pm$0.70$)\times10^{11}$    & \pz 5.7$\pm$0.5 &    $-17\pm$1  \\
K2    & \pz   578$\pm$12\pz  & 0.320$\pm$0.008     & \pp  $179\pm$1   & 0.122$\pm$0.004     & (9.79$\pm$0.85$)\times10^{10}$    & \pz 8.6$\pm$2.0 & \pp $84\pm$5  \\
K3    & \pz\pz 79$\pm$12\pz  & \pz0.79$\pm$0.03\pz &     $-161\pm$2   & 0.231$\pm$0.012     & (3.64$\pm$0.93$)\times10^{9\pz}$  &     \ldots      &   \ldots      \\
K4    & \pz   111$\pm$13\pz  & \pz1.26$\pm$0.04\pz &     $-170\pm$1   & 0.352$\pm$0.018     & (2.22$\pm$0.49$)\times10^{9\pz}$  &     \ldots      &   \ldots      \\
K5    & \pz\pz 70$\pm$13\pz  & \pz1.66$\pm$0.05\pz &     $-173\pm$2   & 0.343$\pm$0.017     & (1.47$\pm$0.42$)\times10^{9\pz}$  &     \ldots      &   \ldots      \\
K6    & \pz   108$\pm$12\pz  & \pz2.39$\pm$0.02\pz &     $-176\pm$1   & 0.247$\pm$0.012     & (4.39$\pm$0.91$)\times10^{9\pz}$  & 20.8$\pm$2.4    & \pp\pz$4\pm$4 \\
K7    & \pz   283$\pm$14\pz  & \pz2.57$\pm$0.04\pz &     $-171\pm$1   & 0.556$\pm$0.056     & (2.26$\pm$0.57$)\times10^{9\pz}$  & 26.2$\pm$6.7    & \pp\pz$8\pm$3 \\
K8    & \pz   214$\pm$15\pz  & \pz3.37$\pm$0.11\pz &     $-168\pm$2   & \pz1.23$\pm$0.12\pz & (3.48$\pm$0.92$)\times10^{8\pz}$  &     \ldots      &   \ldots      \\
%K9   & \pz   284$\pm$17\pz  & 5.13\pz$\pm$0.28\pz &     $-174\pm$3   & 3.03\pz$\pm$0.30\pz & (7.6\pz$\pm$2.0$)\times10^{7\pz}$ &     \ldots      &   \ldots      \\
  \hline
\multicolumn{8}{c}{VLBA-BU-BLAZAR ground array at 43~GHz} \\
  \hline
Core  &      1575$\pm$72\pz  &      \ldots         &   \ldots         & 0.026$\pm$0.005     & (1.48$\pm$0.63)$\times10^{12}$    & \pz 1.7$\pm$0.2 & \pp   $37\pm$6 \\
Q1    &      1373$\pm$65\pz  & 0.091$\pm$0.005     & \pp $177\pm$8    & 0.048$\pm$0.005     & (3.92$\pm$1.00)$\times10^{11}$    & \pz 2.0$\pm$0.3 & \pp\pz $7\pm$4 \\
Q2    & \pz   786$\pm$32\pz  & 0.262$\pm$0.005     & \pp $173\pm$1    & 0.077$\pm$0.007     & (8.68$\pm$1.93)$\times10^{10}$    & \pz 5.4$\pm$0.1 &      $-18\pm$2 \\
Q3    & \pz   402$\pm$18\pz  & 0.378$\pm$0.008     & \pp $177\pm$1    & 0.105$\pm$0.009     & (2.40$\pm$0.52)$\times10^{10}$    & \pz 3.1$\pm$0.8 &      $-29\pm$4 \\
Q4    & \pz\pz 50$\pm$10\pz  & \pz0.92$\pm$0.07\pz &    $-165\pm$5    & 0.239$\pm$0.012     & (5.79$\pm$1.74)$\times10^{8\pz}$  &     \ldots      &   \ldots       \\
Q5    & \pz   100$\pm$11\pz  & \pz1.51$\pm$0.08\pz &    $-170\pm$3    & 0.382$\pm$0.019     & (4.52$\pm$0.95)$\times10^{8\pz}$  &     \ldots      &   \ldots       \\
Q6    & \pz   209$\pm$13\pz  & \pz2.57$\pm$0.09\pz &    $-173\pm$2    & 0.562$\pm$0.028     & (4.34$\pm$0.70)$\times10^{8\pz}$  &     \ldots      &   \ldots       \\
%Q7   & \pz   138$\pm$16\pz  & 3.9\pz\pz$\pm$0.4\pz\pz &$-172\pm$6    & 1.82\pz$\pm$0.09\pz & (2.74$\pm$0.59)$\times10^{7\pz}$  &     \ldots      &   \ldots       \\
  \hline
\multicolumn{8}{c}{Ground array 15~GHz} \\
  \hline
Core  &      2705$\pm$124    &         \ldots      &   \ldots         & $<$0.06             & $>$7.9$\times10^{12}$             & \pz 2.0$\pm$0.1 &      $-20\pm$1 \\
U2    &      1193$\pm$60\pz  & 0.266$\pm$0.019     & \pp $177\pm$5    & $<$0.06             & $>$3.5$\times10^{12}$             & \pz 1.5$\pm$0.1 &      $-30\pm$1 \\
U3    & \pz   266$\pm$23\pz  & 0.982$\pm$0.025     &    $-166\pm$2    & 0.523$\pm$0.026     & (5.03$\pm$0.94)$\times10^{9\pz}$  & \pz 5.5$\pm$1.7 & \pp   $67\pm$3 \\
U4    & \pz\pz 84$\pm$13\pz  & \pz1.53$\pm$0.02\pz &    $-171\pm$1    & 0.235$\pm$0.012     & (7.87$\pm$2.00)$\times10^{9\pz}$  &     \ldots      &   \ldots       \\
U5    & \pz   462$\pm$32\pz  & \pz2.43$\pm$0.02\pz &    $-173\pm$1    & 0.503$\pm$0.025     & (9.44$\pm$1.59)$\times10^{9\pz}$  &    18.9$\pm$4.0 & \pp\pz $5\pm$1 \\
U6    & \pz   332$\pm$27\pz  & \pz3.15$\pm$0.04\pz &    $-167\pm$1    & \pz1.03$\pm$0.05\pz & (1.62$\pm$0.29)$\times10^{9\pz}$  &    27.0$\pm$18  & \pp   $16\pm$8 \\
U7    & \pz   391$\pm$33\pz  & \pz4.96$\pm$0.12\pz &    $-173\pm$2    & \pz2.84$\pm$0.14\pz & (2.52$\pm$0.46)$\times10^{8\pz}$  &    \ldots       &    \ldots      \\
  \hline
\end{tabular}
\end{center} {\bf Notes.} Tabulated data correspond to: component's label; flux density; distance and position angle from the core; size; observed brightness temperature, degree of linear polarization; and electric vector position angle, \emph{uncorrected} for Faraday rotation (see Sec.~\ref{Sec:pol}).
\end{table*}

\subsection{Stationary components in the innermost 0.5~mas region}
\label{sec:sta}

  Previous high angular resolution monitoring programs of BL~Lac systematically show the presence of two stationary features close to the core \citep{2003MNRAS.341..405S,2005AJ....130.1418J,2005ApJ...623...79M}. In particular, \cite{2005AJ....130.1418J} report, through an analysis of a sequence of 17 bimonthly VLBA observations at 43~GHz, two stationary components, labeled A1 and A2, located at a mean distance from the core (position angle) of 0.10~mas and 0.29~mas ($-160^{\circ}$ and $-159^{\circ}$), respectively. These can be associated with components C2 and C3 in \cite{2005ApJ...623...79M}, respectively, obtained from independent 43~GHz VLBA observations.

  More recently, \cite{2014ApJ...787..151C,2015ApJ...803....3C} present an analysis of more than a decade of 15~GHz VLBA observations of BL~Lac from the MOJAVE monitoring program. These observations confirm the existence of a stationary component, labeled C7 by these authors, located at a mean distance from the core of 0.26~mas and at a position angle of $-166${\mbox{$.\!\!^\circ$}}6, in agreement with component A2 reported by \cite{2005AJ....130.1418J}. 
  
  This is also corroborated by our 15~GHz and 43~GHz observations (see Table~\ref{Tb:modfit} and Figs.~\ref{Fig:U-band} and \ref{Fig:Q-band}), in which components U2 and Q2 would correspond to previously identified components A2 and C7, and component Q1 to A1.
  
  Our measured position angles for the two stationary features (Q1 and Q2/U2) are slightly offset to the east by $\sim$20$^{\circ}$ with respect to the main values published by \cite{2005AJ....130.1418J} and \cite{2014ApJ...787..151C}. This may be associated with the jet precession reported by \cite{2003MNRAS.341..405S} and \cite{2005ApJ...623...79M}, leading to a swing in the position angle of the innermost components. A similar variation in the position angle of component U2 is seen in MOJAVE observations by \cite{2014ApJ...787..151C}.

\subsection{Evidence for emission upstream of the core}

  Model fitting of the innermost structure in our \emph{RadioAstron} observations, listed in Table~\ref{Tb:modfit} and plotted in Fig.~\ref{Fig:RA_BLLac}, shows two close components near the core region, as well as two other components within the innermost 0.5 mas region. The fitted circular Gaussian components provide an accurate representation of the jet emission, yielding a residual map (uniform weighting) with a rms of 1.4 mJy/beam and minimum and maximum residuals of -45 mJy/beam and 53 mJy/beam, respectively. Allowing for elliptical Gaussian components provides very similar fitted values as those listed in Table~\ref{Tb:modfit} with no significant improvement in the residuals.
  
  We can tentatively identify components K1 and K2 with the previously discussed stationary features Q1 and Q2 (U2). However identification of the two-component structure in the core area requires a more detailed analysis of the evolution of the innermost structure of the jet.
  
\begin{figure}[t]
\epsscale{1.12}
\plotone{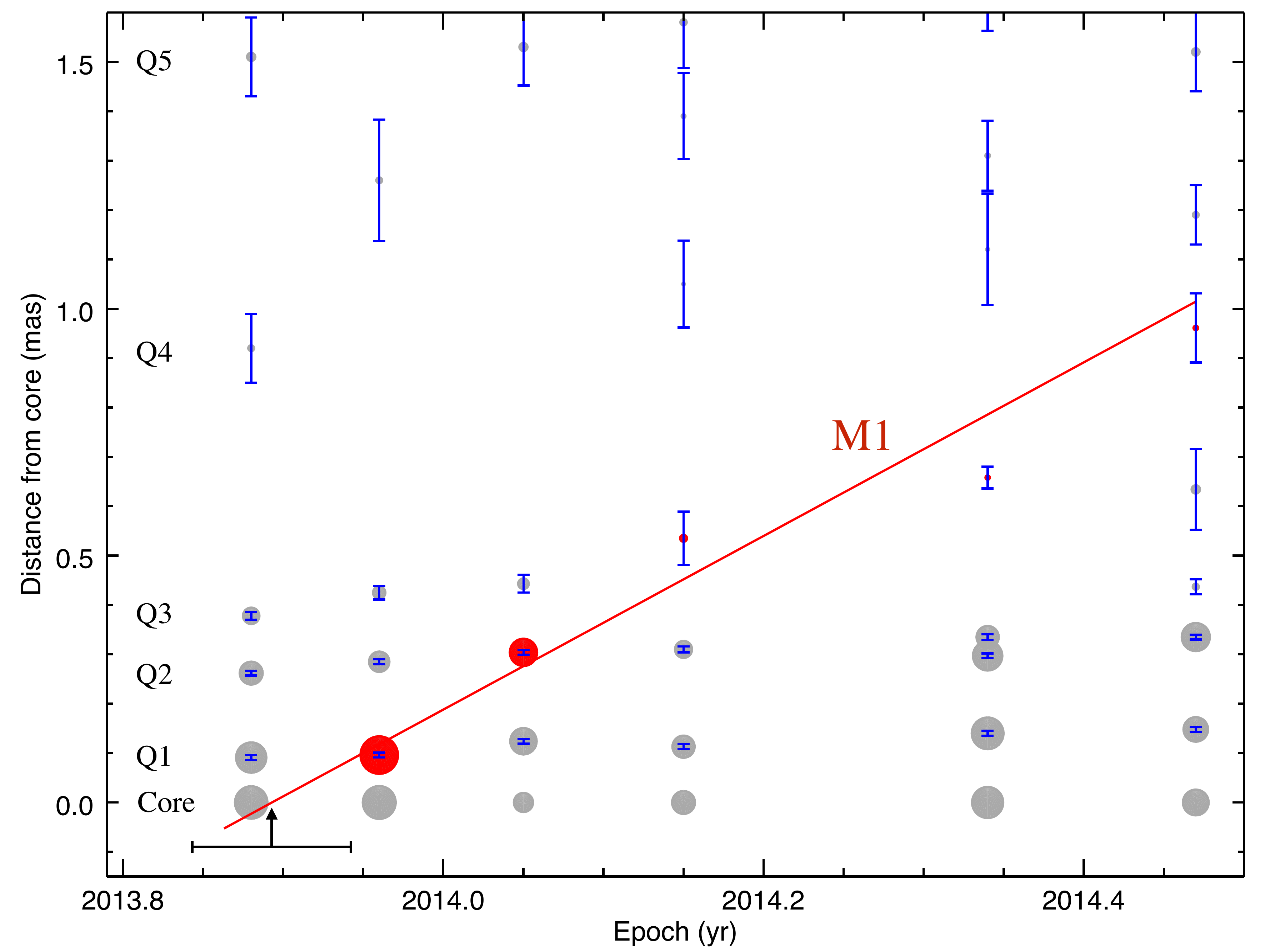}
\caption{VLBA-BU-BLAZAR model fits covering November 2013 through June 2014. Symbol size is proportional to the square root of component's flux density. Marked in red is a new superluminal component, M1, with a proper motion of 7.9$\pm0.3\,c$, ejected from the core in 2013.89$\pm$0.05 (marked with an arrow and its respective error bar).}
\label{Fig:Q-fits}
\end{figure}

  For this we have performed model fitting of the 43~GHz VLBA images from the VLBA-BU-BLAZAR program extending our analysis to cover the period between 2013 December and 2014 June, comprising a total of five more epochs with a cadence of roughly one month. The obtained model fits are listed in Table~\ref{Tb:43_mf}, and Fig.~\ref{Fig:Q-fits} plots the component's distance from the core versus time. Stationary components Q1 and Q2 are detected at all epochs. Analysis of their flux densities show that Q1 becomes unusually bright (2.1 Jy) on 2013 December 16, followed by a similar increase in flux density of Q2 (1.1 Jy) on 2014 January 19. This could be associated with the passing of a new component, M1, through the standing features, leading to a brief increase in their flux densities \citep[e.g.,][]{1997ApJ...482L..33G}. Component M1 is identified in the last two epochs as a weak knot beyond 0.5 mas from the core. The estimated apparent speed for this new component is 7.9$\pm0.3\,c$ (1.76$\pm$0.06 mas/yr), in agreement with values previously found in BL~Lac \citep{2005AJ....130.1418J}, giving an ejection date of 2013.89$\pm$0.05, or 2013 November 23 ($\pm$18 days), in coincidence with our \emph{RadioAstron} observations within the errors.
  
  Considering the measured proper motion of the new component and the estimated time of crossing through the 43~GHz core, we can estimate that component M1 should be placed $\sim$50~$\mu$as \emph{upstream} of the core in our \emph{RadioAstron} image, or slightly smaller if we account for some initial acceleration. Based on this, we consider that the core of the jet in our \emph{RadioAstron} observations corresponds to the unresolved component labeled ``Core'', upstream of which component ``Up'' would correspond to component M1 identified also at 43~GHz. Evidence for emission upstream the core is also found by looking at the 43~GHz polarization image (see Fig.~\ref{Fig:Q-band}), which shows polarized emission to the north with a different orientation of the polarization vectors than the remaining core area.

\begin{figure*}[t]
  \centering
  \includegraphics[width=0.85\textwidth]{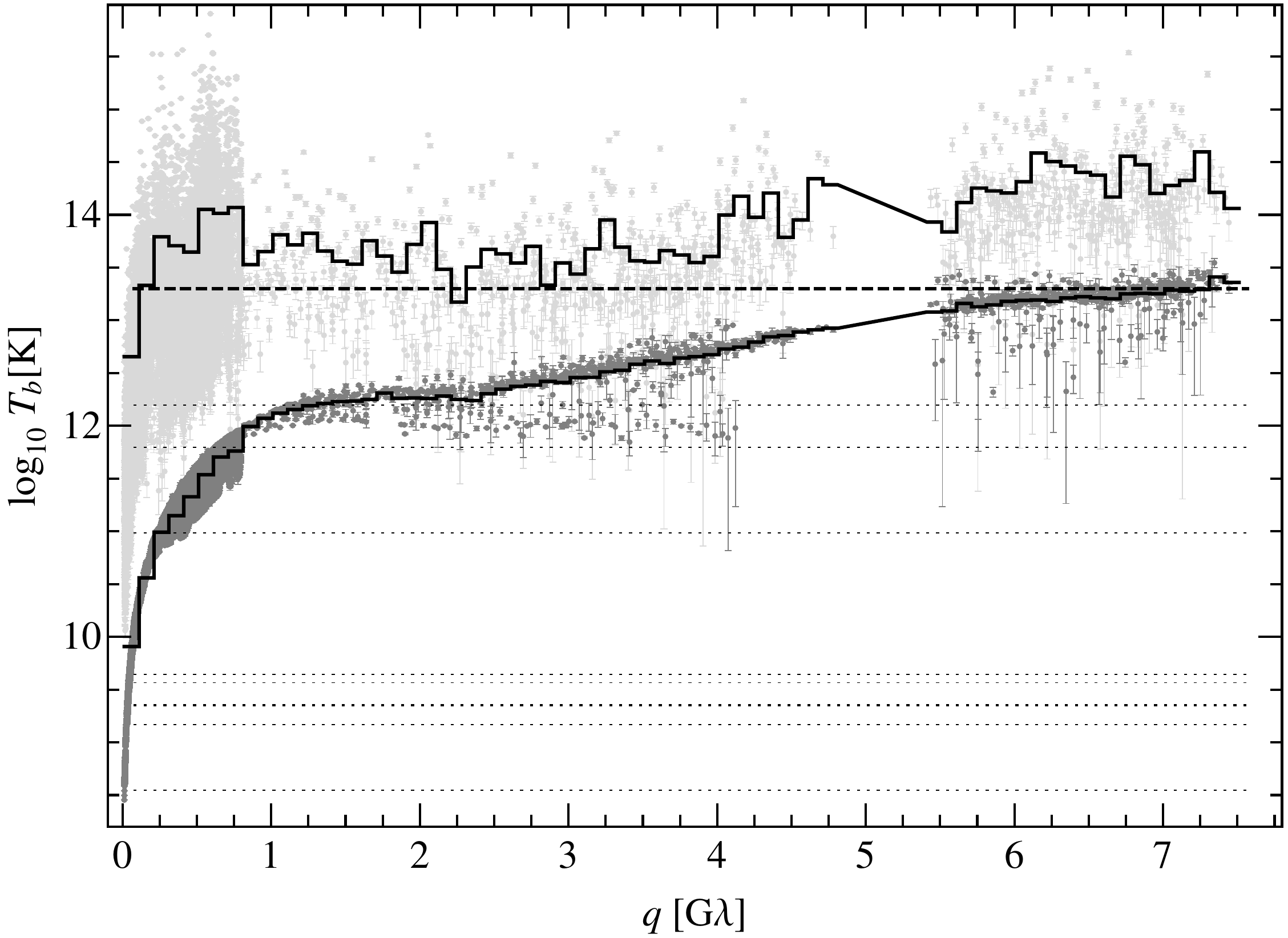}
  \caption{Visibility-based estimates of the observed brightness temperature in BL~Lac, calculated from the RadioAstron data at 22\,GHz. Data points represent the minimum (dark gray), $T_\mathrm{b,min}$, and maximum (light gray), $T_\mathrm{b,max}$, brightness temperatures derived from individual visibilities measured at different {\em uv} distances, $q$ \citep{2015A&A...574A..84L}. The two-dimensional distribution of $T_\mathrm{b,min}$ is also presented in Fig.~\ref{Fig:uvplot} with a color wedge. The histograms show the respective brightness temperatures averaged over bins of 0.1\,G$\lambda$ in size. The thick dashed line indicates the lower limit on brightness temperature, $T_\mathrm{b,mod} > 2\times 10^{13}$\,K, derived for the core component of the source structure described in Table~\ref{Tb:modfit}. The dotted lines indicate the respective brightness temperatures of the rest of the model components. The minimum brightness temperature, $T_\mathrm{b,min} = 1.5 \times 10^{13}$\,K, is constrained by the visibilities with $q>5.3$\,G$\lambda$. The estimated maximum brightness temperature $T_\mathrm{b,max} = 3.1 \times 10^{14}$\,K is calculated from the data on the same baselines.}
\label{Fig:bllac-tb} 
\end{figure*}

  We note that the appearance of new superluminal components \emph{upstream} of the core in BL~Lac has been previously reported by \cite{Marscher:2008ii}, associated with a multi-wavelength outburst. Similarly, the radio, optical, and $\gamma$-ray light curves (see the VLBA-BU-BLAZAR web page\footnote{\url{https://www.bu.edu/blazars/VLBA\_GLAST/bllac.html}}) show a flare at the end of 2013, close to our \emph{RadioAstron} observations.
  
\subsection{Brightness temperature}
\label{Sec:Tb}

  Table \ref{Tb:modfit} also lists the observed (i.e., not corrected for redshift or Doppler boosting) brightness temperatures, estimated from the model-fitted circular Gaussian components as $T_\text{b}=1.36\times10^9S\lambda^2/\theta^2$\,K, where $S$ (in Jansky) is the total flux density, $\theta$ (mas) the size, and $\lambda$ (cm) the observing wavelength \citep[e.g.,][]{2005AJ....130.2473K}. Model-fitted data of the \emph{RadioAstron} visibilities yield a brightness temperature of (1.56$\pm$0.26$)\times10^{12}$ K for the component upstream the core, ``Up'', while for the unresolved core component, ``Core'', we obtain a lower limit of $T_\text{b}>$2.0$\times10^{13}$\,K.

  The observed brightness temperature in jets is mostly affected by the transverse dimension of the flow and it may differ systematically from estimates obtained on the basis of representing the jet structure with Gaussian components \citep{2015A&A...574A..84L}. In this case, constraints on the jet brightness temperature can also be found directly from visibility amplitudes and their errors, providing the minimum brightness temperature, $T_\mathrm{b,min}$, and an estimate of the formal maximum brightness temperature, $T_\mathrm{b,max}$ that can be obtained under condition that the structural detail sampled by the given visibility is resolved. For further details we refer the reader to \cite{2015A&A...574A..84L}.

  These two estimates are compared in Fig.~\ref{Fig:bllac-tb} with the brightness temperatures calculated from the Gaussian components described in Table~\ref{Tb:modfit}. One can see that the observed brightness temperature of the most compact structures in BL~Lac, constrained by baselines longer than $5.3\,\mathrm{G}\lambda$, must indeed exceed $2\!\times\!10^{13}$\,K and can reach as high as $\sim\!3\!\times\! 10^{14}$\,K. As follows from Fig.~\ref{Fig:uvplot}, these visibilities correspond to the structural scales of 30--40\,$\mu$as oriented along position angles of $25^{\circ}-30^{\circ}$. These values are indeed close to the width of the inner jet and the normal to its direction.
  
  The observed, $T_\text{b,obs}$, and intrinsic, $T_\text{b,int}$, brightness temperatures are related by 
  $$T_\text{b,obs}=\delta(1+z)^{-1}T_\text{b,int}$$
where $\delta=(1-\beta^2)^{1/2}(1-\beta\cos\phi)^{-1}$ is the Doppler factor, $\beta$ is the jet bulk velocity in units of the speed of light, $\phi$ is the jet viewing angle, and $z$ is the redshift of the source. Variability arguments \citep{2005AJ....130.1418J,2009A&A...494..527H} and kinematical analyses \citep{2015ApJ...803....3C} yield a remarkably consistent value of $\delta=7.2$, from which we estimate a lower limit of the intrinsic brightness temperature in the core component of our \emph{RadioAstron} observations of $T_\text{b,int}>2.9\times10^{12}$\,K.%, and can reach values as high as $4.5\times10^{13}$\,K for the $T_\mathrm{b,lim}$ estimates on baselines longer than $5.3\,\mathrm{G}\lambda$.

  It is commonly considered that inverse Compton losses limit the intrinsic brightness temperature for incoherent synchrotron sources, such as AGN, to about $10^{12}$~K \citep{1969ApJ...155L..71K}. In case of a strong flare, the ``Compton catastrophe'' is calculated to take about one day to drive the brightness temperature below $10^{12}$~K. Moreover, \cite{1994ApJ...426...51R} has argued that for sources near equipartition of energy between the magnetic field and radiating particles a more accurate upper value for the intrinsic brightness temperature is about $10^{11}$~K \citep[see also][]{1999ApJ...511..112L,2009A&A...494..527H}, which is often called the equipartition brightness temperature.
  
  Our estimated lower limit for the intrinsic brightness temperature of the core in the \emph{RadioAstron} image of $T_\text{b,int}>2.9\times10^{12}$\,K is therefore more than an order of magnitude larger than the equipartition brightness temperature limit established by \cite{1994ApJ...426...51R}, and at least several times larger than the limit established by inverse Compton cooling. This suggests that the jet in BL~Lac is not in equipartition, as may be expected in case the source is flaring during the ejection of component M1 detected at 43~GHz, and rises the possibility that we are significantly underestimating its Doppler factor.
  
  We also note that if our estimate of the maximum brightness temperature is closer to actual values, it would imply $T_\mathrm{b,int}\sim5\times10^{13}$\,K. This is difficult to reconcile with current incoherent synchrotron emission models from relativistic electrons, requiring alternative models such as emission from relativistic protons \citep{1967Natur.216..461J}, or coherent emission \citep{Benford:2000ic} -- see also \cite{2002PASA...19...77K} and references therein.
   
\section{Polarization and Faraday rotation analysis}
\label{Sec:pol}

\begin{figure}[t]
\epsscale{1.2}
\plotone{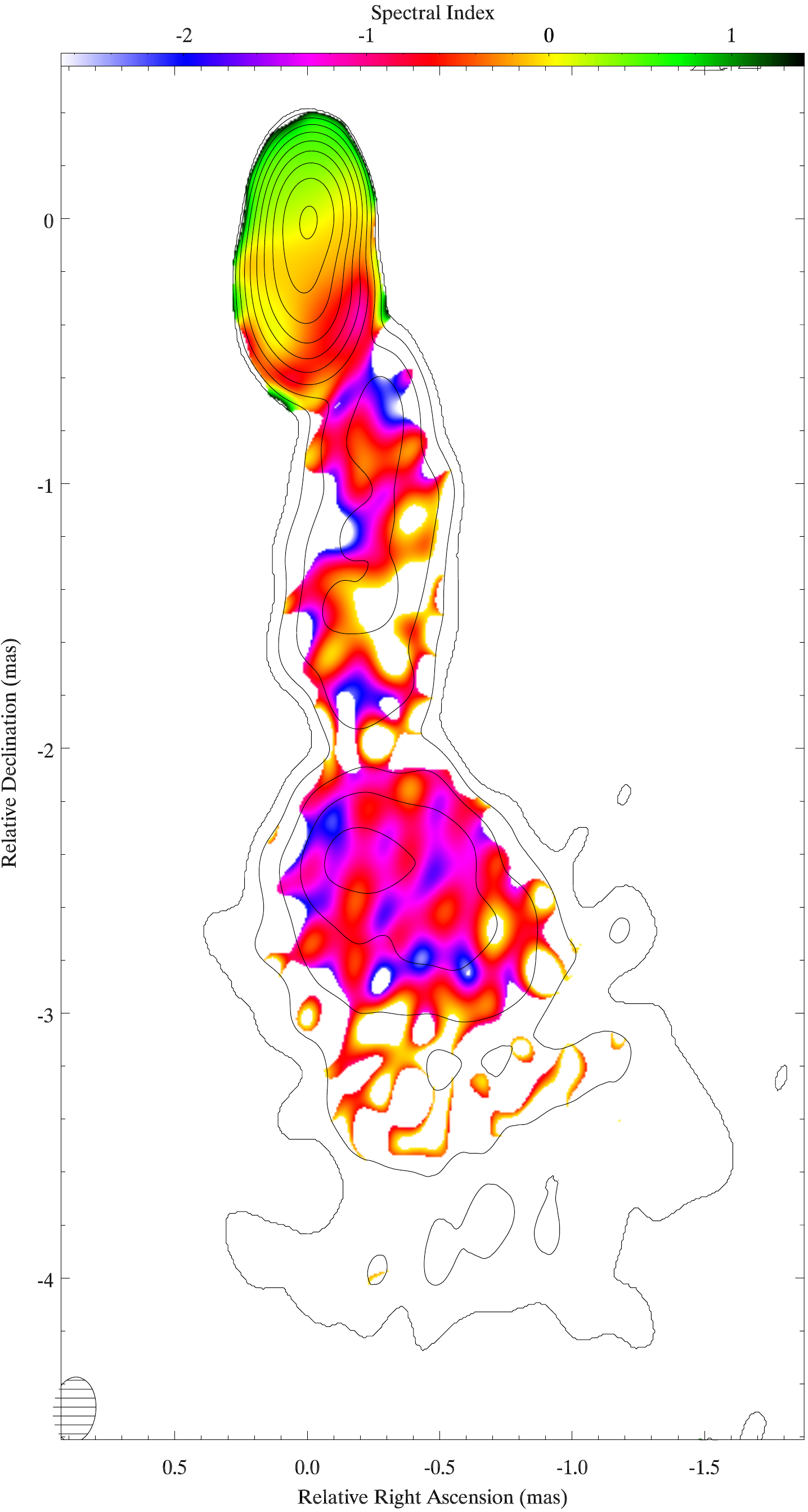}
\caption{Spectral index image between the 22~GHz \emph{RadioAstron} and 43~GHz ground-array total intensity images. Contours show the total intensity image from the \emph{RadioAstron} observations convolved with the 43~GHz beam.}
\label{Fig:Q-K_spnd}
\end{figure}

 Our polarimetric observations with the ground arrays at 15 and 43~GHz and the space VLBI \emph{RadioAstron} observations at 22~GHz can be combined to obtain a rotation measure (RM) image of BL~Lac. Data at 43 and 22~GHz were first tapered and convolved with a common restoring beam to match the 15~GHz resolution.

  Combination of the images at all three frequencies requires also proper registering. Due to the difficulties in finding compact, optically thin components that could be matched across images, we have used for the image alignment a method based on a cross-correlation analysis of the total intensity images \citep{Walker:2000gb,Croke:2008bn,Hovatta:2012jv}. Only optically thin regions have been considered in the cross-correlation analysis to avoid the shifts in the core position due to opacity \citep[e.g.,][]{Lobanov:1998vr,Kovalev:2008hm} that could influence our results. We obtain a shift to the south, in the direction of the jet, of 0.021 mas and 0.063 mas for the alignment of the 22 and 15~GHz images, respectively, with respect to the one at 43~GHz.

  Pixels in the images for which polarization was not detected at all three frequencies simultaneously were blanked. The rotation measure (RM) map is computed by performing a $\lambda^2$ fit to the wavelength dependence of the EVPAs at each pixel, blanking pixels with a poor fit based on a $\chi^2$ criterion. Due to the $n\pi$ ambiguity in the EVPAs, we have developed an IDL routine that searches for possible $n\pi$ rotations, finding that no wraps higher than $\pm\pi$ were required to fit the data.

\begin{figure}[t]
\epsscale{1.2}
\plotone{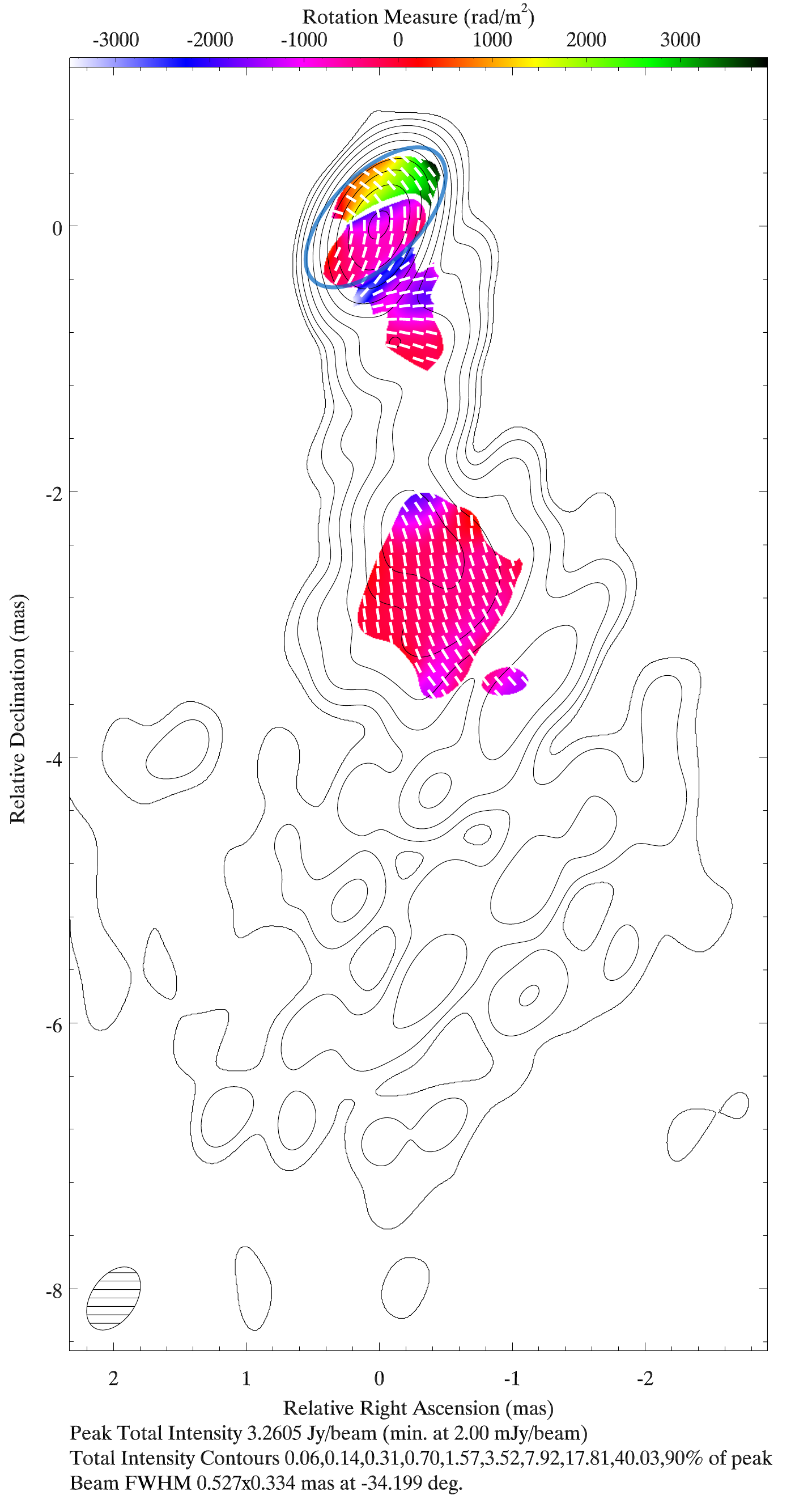}
\caption{Rotation measure map combining the ground-array images at 15 and 43~GHz with the \emph{RadioAstron} image at 22~GHz. Contours show the 15~GHz image, colors indicate the rotation measure, and bars plot the \emph{Faraday-corrected} EVPAs. The light blue ellipse delimits the core region within which the 2D histogram images (see Fig.~\ref{Fig:H2D}) have been computed.}
\label{Fig:RM}
\end{figure}

\begin{figure}[t]
\epsscale{1.2}
\plotone{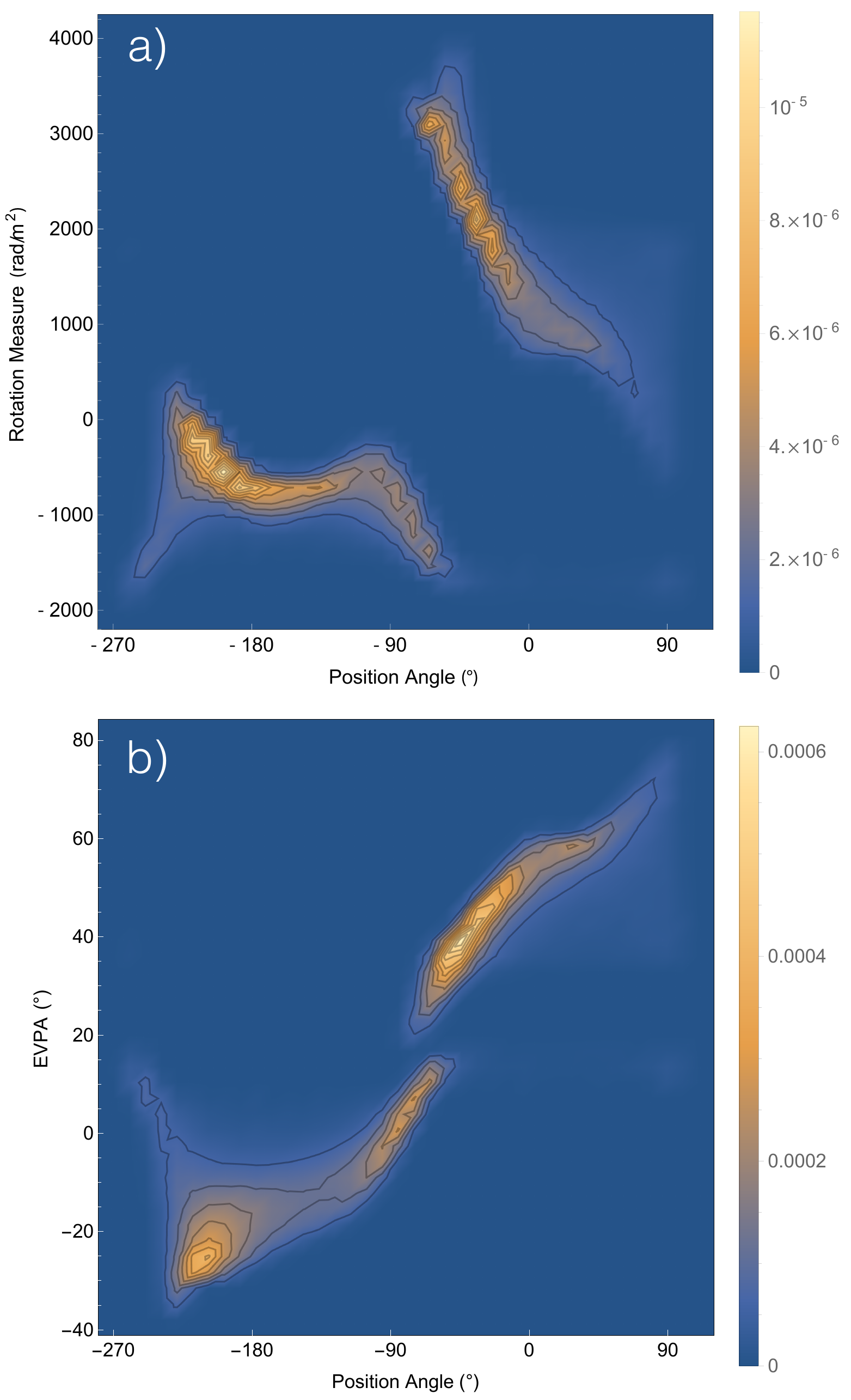}
\caption{Probability density functions of the two-dimensional histograms of the rotation measure (\emph{a, top}) and Faraday corrected EVPA (\emph{b, bottom}) for the core area indicated by the blue ellipse in Fig.~\ref{Fig:RM}.}
\label{Fig:H2D}
\end{figure}
  
  When performing the RM analysis of the core area, it is also important to pay special attention to possible $\pi/2$ rotations due to opacity \citep[e.g.,][]{1994A&A...292...33G,2001MNRAS.320L..49G,Porth:2011ev}. We have checked for these by first computing the spectral index maps between each pair of frequencies. Figure \ref{Fig:Q-K_spnd} shows the spectral index map between the 22~GHz \emph{RadioAstron} and 43~GHz VLBA images. This reveals an optically thick region at 22~GHz (and therefore also at 15~GHz) at the upstream end of the jet, near the core. This optically thick region is accounted for when computing the RM map by rotating the EVPAs at 22 and 15~GHz by $\pi/2$. The resulting rotation measure map is shown in Fig.~\ref{Fig:RM}.

\subsection{Evidence for a helical magnetic field}

  The rotation measure and RM-corrected EVPAs ($\chi_0$) in the core area (delimited by the blue ellipse in Fig.~\ref{Fig:RM}) exhibit a clear point symmetry around its centroid. To better analyze this structure Fig.~\ref{Fig:H2D} displays the probability distribution function of the two-dimensional histogram for the RM and $\chi_0$ as a function of the position angle of the pixels with respect to the centroid of the core, measured counterclockwise from north.

  By inspecting Figs.~\ref{Fig:RM} and \ref{Fig:H2D}a we note a gradient in RM with position angle from the centroid of the core, with positive RM values in the area upstream of the centroid, and negative downstream, in the direction of the jet. The largest RM values, of the order of 3000 rad m$^{-2}$, are found in the area northeast of the centroid (with a position angle of $\sim-45^{\circ}$); smaller values are obtained as the position angle increases, reaching values of $\sim$1000 rad m$^{-2}$ north-west of the centroid. Downstream of the centroid the RM continues this trend, reaching values of $\sim$-900 rad m$^{-2}$ in the direction of the jet.

  Similarly, Figs.~\ref{Fig:RM} and \ref{Fig:H2D}b show a progressive rotation in $\chi_0$ with position angle from the centroid of the core. Counterclockwise from east, $\chi_0$ rotate continuously from $\sim-25^{\circ}$ to $\sim60^{\circ}$. On top of this, the two-dimensional histogram shows a concentration of $\chi_0$ values between approximately 30$^{\circ}$ and 50$^{\circ}$ in the area upstream of the centroid, while downstream these concentrate in values at $\sim-25^{\circ}$ and 0$^{\circ}$.

  A similar dependence of the RM and $\chi_0$ with polar angle was found by \cite{Zamaninasab:2013bn} in 3C~454.3,  interpreted by these authors as the result of a helical magnetic field. Indeed, gradients in RM across the jet width are expected to arise in the case of helical magnetic fields \citep{Laing:1981bx}, as previously reported in a number of sources \citep[e.g.,][]{Asada:2002to,OSullivan:2009bx,Hovatta:2012jv}.

\begin{figure}
\epsscale{1.15}
\plotone{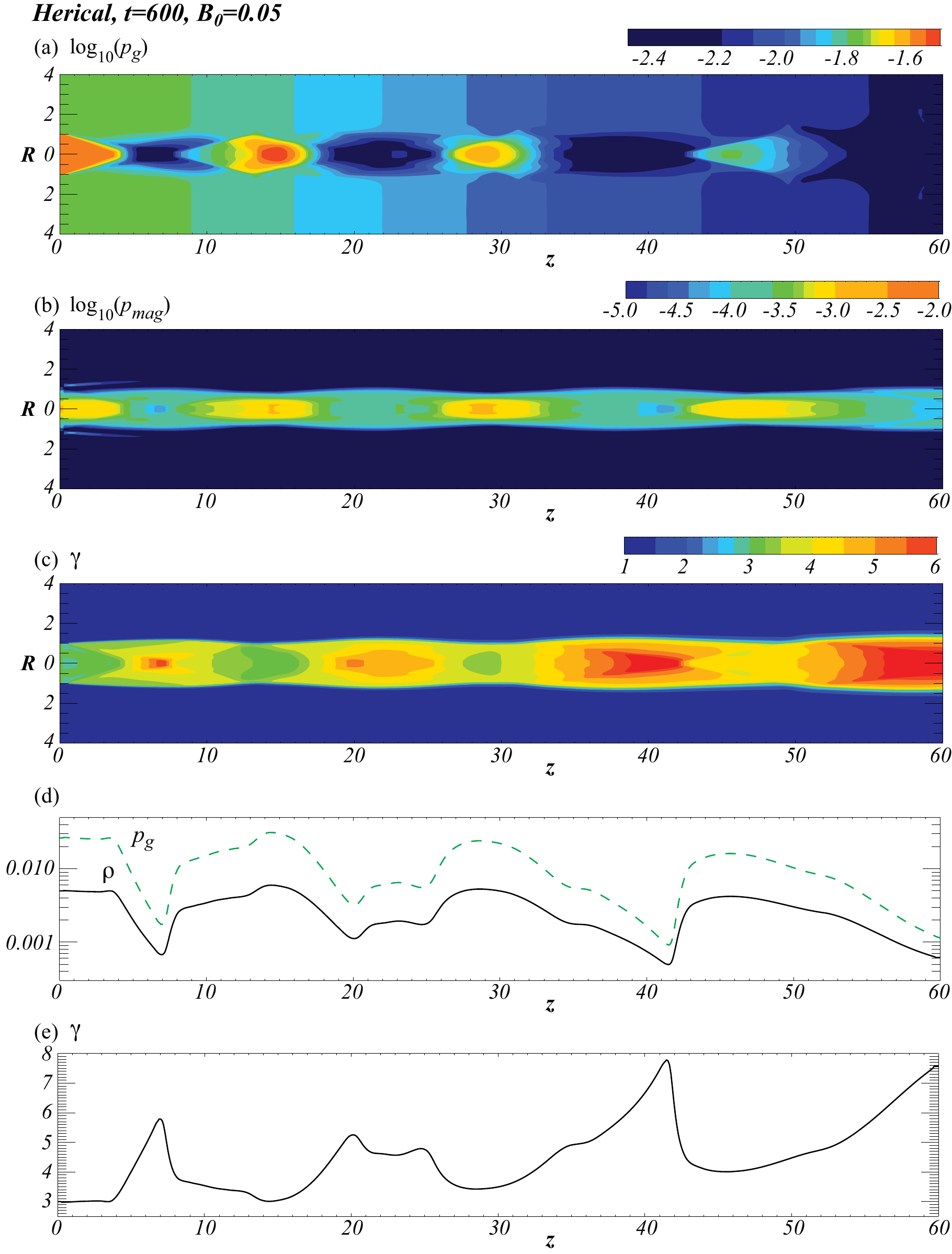}
\caption{Relativistic magnetohydrodynamic simulations for an over-pressured jet propagating from left to right with a helical magnetic field at $t_\text{s}=600$, where $t_\text{s}$ is in units of $R_\text{j}/c$. {\it Upper panels:} 2D plots of: ({\it a}) the gas pressure density, ({\it b}) the magnetic pressure ($p_\text{mag}=B^2/2$), and ({\it c}) the Lorentz factor. {\it Lower panels:} 1D profiles along the jet axis ($R=0$) of: ({\it d}) the rest-mass density (solid) and the gas pressure (green dashed), and ({\it e}) the Lorentz factor.}
\label{Fig:2D_RMHD}
\end{figure}

\begin{figure}[t]
\epsscale{1.17}
\plotone{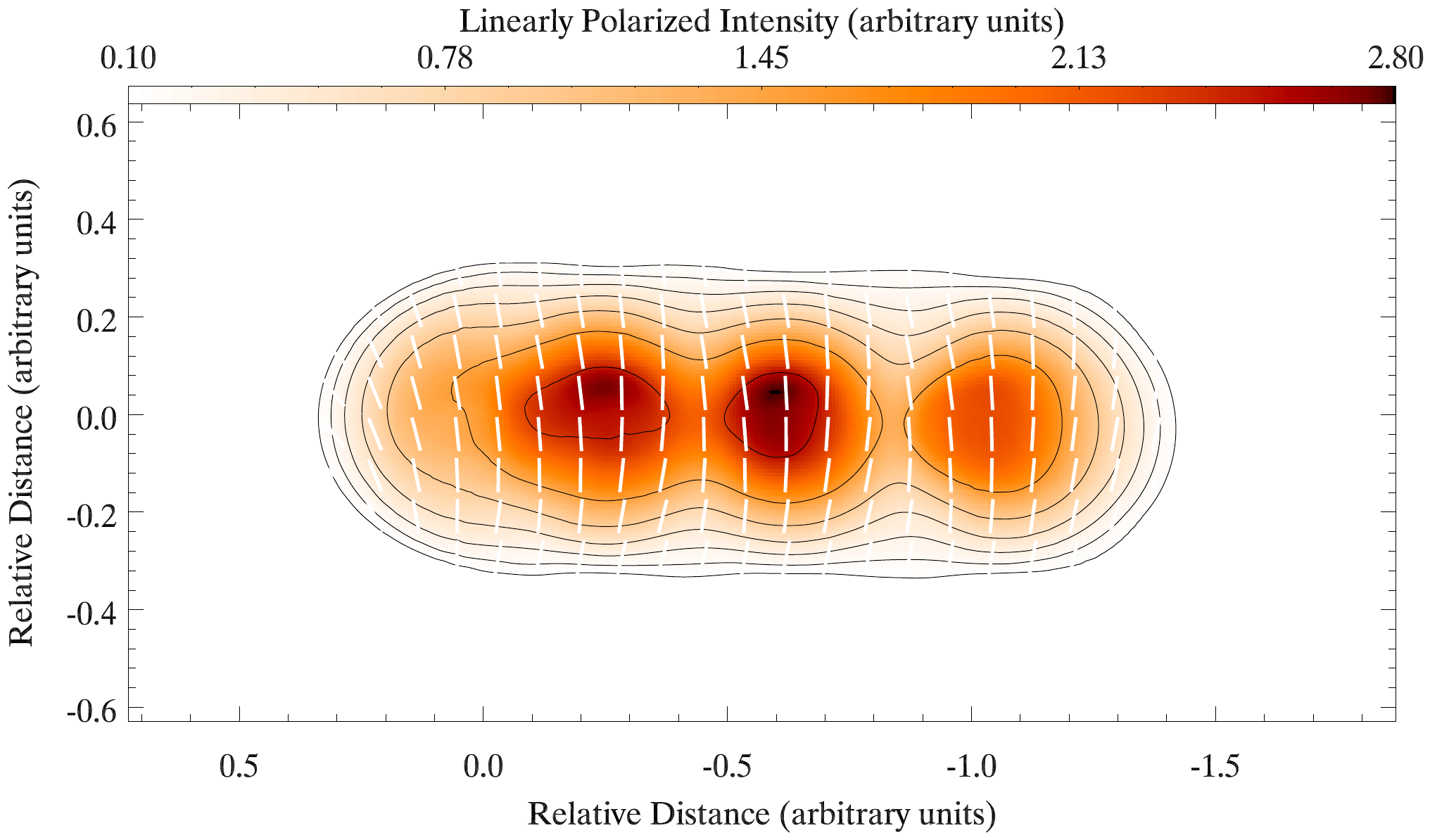}
\caption{Simulated total intensity (contours, in logarithmic scale), linearly polarized intensity (colors), and EVPAs (bars) obtained by computing the (optically thin) radio continuum synchrotron emission at a viewing angle of 10$^{\circ}$ using the RMHD model shown in Fig.~\ref{Fig:2D_RMHD} as input.}
\label{Fig:sim_10deg}
\end{figure}

  Relativistic magnetohydrodynamic (RMHD) simulations have been used to study the rotation measure and polarization distribution based on self-consistent models for jet formation and propagation in the presence of large scale helical magnetic fields \citep{Broderick:2010dy,Porth:2011ev}. These simulations reproduce the expected gradients in RM across the jet width due to the toroidal component of the helical magnetic field, and provide also detailed insights regarding the polarization structure throughout the jet, which depends strongly on the helical magnetic field pitch angle, jet viewing angle, Lorentz factor, and opacity. 
  
  As discussed in \cite{Zamaninasab:2013bn}, \cite{Broderick:2010dy}, and \cite{Porth:2011ev}, a large-scale helical magnetic field would lead to similar point symmetric structures around the centroid of the core of both RM and $\chi_0$ as found in Figs.~\ref{Fig:RM} and \ref{Fig:H2D}. This suggests that the core region in BL~Lac is threaded by a large-scale helical magnetic field. A more detailed comparison between our observations and specific RMHD simulations using the estimated parameters for BL~Lac is underway and will be published elsewhere.

\subsection{Pattern of recollimation shocks}

  The location of the stationary feature $\sim$0.26~mas from the core (see Sect.~\ref{sec:sta}) marks a clear transition in the RM and polarization vectors between the core area and the remainder of the jet in BL~Lac. Figure \ref{Fig:RM} shows a localized region of enhanced RM, reaching values of the order of $-2200\pm$300 rad m$^{-2}$, and a Faraday rotation corrected EVPA of $-40\pm8^{\circ}$. $\chi_0$ therefore becomes perpendicular to the local jet direction, suggesting a dominant component of the magnetic field that is aligned with the jet. Downstream of this location, $\chi_0$ rotates so that the magnetic field remains predominantly aligned with the local jet direction up to a distance of $\sim$1~mas from the core.

  Further downstream polarization is detected again in a region at a distance of $\sim$3~mas from the core, corresponding to the location of components K6 and K7 (see Fig.~\ref{Fig:RA_BLLac}). This area has a mean RM and $\chi_0$ of $-320\pm140$ rad m$^{-2}$ and 14$\pm$6$^{\circ}$, respectively. The polarization vectors are approximately parallel to the jet in this area, thereby suggesting that the magnetic field is predominantly perpendicular to the jet. This would be in agreement with a helical magnetic field in which the toroidal component dominates over the poloidal one, although other scenarios, like a plane perpendicular shock, cannot be ruled out.

  Considering that the polarization properties in the core area and components K6 and K7 suggest that the jet in BL~Lac is threaded by a helical magnetic field, the different polarization structure associated with the stationary feature at $\sim$0.26 mas from the core suggests that it may correspond to a rather particular jet feature. This would be in agreement with claims by \cite{2015ApJ...803....3C}, and references therein, in which these authors conclude that this stationary feature corresponds to a recollimation shock, downstream of which new superluminal components appear due to the propagation of Alfv\'en waves triggered by the swing in its position, in a similar way as exciting a wave on a whip by shaking the handle.

  Previous multiwavelength observations of BL~Lac suggest that the core may also correspond to a recollimation shock \citep{Marscher:2008ii}. In that case we can hypothesize that the upstream component found in our \emph{RadioAstron} observation (see Fig.~\ref{Fig:RA_BLLac} and Table~\ref{Tb:modfit}) is located at the jet apex, so that the distance at which the first recollimation shock takes place, associated with the core, would be $\sim$40~$\mu$as. If so, the jet in BL~Lac would contain a set of three recollimation shocks, spaced at progressively larger distances of approximately 40, 100, and 250~$\mu$as.

  To investigate such a pattern of recollimation shocks as possibly observed in BL~Lac, we have performed two-dimensional special relativistic magnetohydrodynamic simulations in cylindrical geometry using the RAISHIN code \citep{2006astro.ph..9004M,Mizuno:2011dm}. The initial set-up follows \cite{1997ApJ...482L..33G} and \cite{2015ApJ...809...38M}, in which a preexisting over-pressured flow is established across the simulation domain. We choose a rest-mass density ratio between the jet and ambient medium of $\eta=\rho_\text{j}/\rho_\text{a} = 5 \times 10^{-3}$ with $\rho_\text{a}=1.0\rho_0$, initial jet Lorentz factor $\gamma_\text{j}=3$, and local Mach number $M_\text{s}=1.69$. The gas pressure in the ambient medium decreases with axial distance from the jet base following $p_{\text{g},\text{a}}(z) = p_{\text{g},0}/[1+(z/z_\text{c})^n]^{m/n}$, where $z_\text{c}=60 R_\text{j}$ is the gas pressure scale height in the axial direction, $R_\text{j}$ is the jet radius, $n=1.5$, $m=2.3$, and $p_{\text{g},0}$ is in units of $\rho_0\,c^2$ \citep[e.g.,][]{1995ApJ...449L..19G,1997ApJ...482L..33G,2009ApJ...696.1142M}. We assume that the jet is initially uniformly over-pressured with $p_{\text{g},\text{j}}=1.5 p_{\text{g},\text{a}}$. We consider a force-free helical magnetic field with a weak magnetization $B_0=0.05$ (in units of $\sqrt{4\pi\rho_0\,c^2}$) and constant magnetic pitch $P_0=R B_z/R_\text{j} B_{\phi}=1/2$, where $B_z$ and $B_{\phi}$ are the poloidal and toroidal magnetic field components, so that smaller $P_0$ refers to increased magnetic helicity.

  Overexpansions and overcontractions caused by inertial overshooting past equilibrium of the jet lead to the formation of a pattern of standing oblique recollimation shocks and rarefactions, whose strength and spacing are governed by the external pressure gradient (see Fig.~\ref{Fig:2D_RMHD}). The jet is accelerated and conically expanded slightly by the gas pressure gradient force, reaching a maximum jet Lorentz factor of $\sim 8$. The jet radius expands from the initial $R_\text{j}=1$ at the jet base to $R_\text{j}=1.5$ at $z=60 R_\text{j}$, yielding an opening angle of $\theta_\text{j} = \arctan (0.5R_\text{j}/60R_\text{j}) \sim 0.48^{\circ}$. The recollimation shocks are located at progressively larger axial distances of $\sim 7 R_\text{j}$, $\sim 20 R_\text{j}$, and $\sim 42R_\text{j}$. The relative distances between the second and third shocks with respect to the first one are $\sim 2.9$ and $\sim 6$, which roughly match those observed in BL~Lac.

  The output from the RMHD simulation is used as input to compute the radio continuum synchrotron emission map shown in Fig.~\ref{Fig:sim_10deg}. Following \cite{1995ApJ...449L..19G}, the internal energy is distributed among the non-thermal electrons using a power law $N(E)dE=N_0 E^{-p} dE$, with $p=2.4$. The emission is computed for a viewing angle of $10^{\circ}$ and at an optically thin observing frequency, integrating the synchrotron transfer equations along the line of sight \citep[e.g.,][]{1997ApJ...482L..33G}. The pattern of recollimation shocks seen in Fig.~\ref{Fig:2D_RMHD} leads to a set of knots in the total and linearly polarized intensity. The EVPAs in the knots are perpendicular to the jet direction, in agreement with the observations of component K2 in BL~Lac. We note however that our RMHD simulations consider an already formed and collimated jet, and therefore do not provide an accurate account of the jet formation region, close to where the other two innermost recollimation shocks are located.
  
  Finally, although our simulations are in general agreement with the pattern of recollimation shocks observed in BL~Lac, further numerical simulations, in progress, are required to better constrain the jet parameters of the model, such as the Lorentz factor, Mach number, gradient in external pressure, magnetic field helicity/strength, and viewing angle.
  
\section{Summary}
\label{Sec:sum}

  \emph{RadioAstron} provides the first true full-polarization capabilities for space VLBI observations on baselines longer than the Earth's diameter, opening the possibility to achieve unprecedentedly high angular resolution in astronomical imaging. In this paper we present the first polarimetric space VLBI observations at 22~GHz, obtained as part of our \emph{RadioAstron} KSP designed to probe the innermost regions of AGN and their magnetic fields in the vicinity of the central black hole.
  
  The jet of BL~Lac was observed in 2013 November 10 at 22~GHz with \emph{RadioAstron} including a ground array of 15 radio telescopes. The instrumental polarization of the space radio telescope was found to remain within 9\% (5\% for LCP), demonstrating its polarization capabilities for \emph{RadioAstron} observations at its highest observing frequency of 22~GHz.
  
  The phasing of a group of ground-based antennas allowed to obtain reliable ground-space fringe detections up to projected baseline distances of 7.9 Earth's diameters in length. Polarization images of BL~Lac are obtained with a maximum angular resolution of 21~$\mu$as, the highest achieved to date.

  Analysis of the 43~GHz data from the VLBA-BU-BLAZAR monitoring program, covering from November 2013 to June 2014, reveals a new component ejected near the epoch of our \emph{RadioAstron} observations, confirmed also by flares in the optical and $\gamma$-ray light curves. This new component appears in the \emph{RadioAstron} image as a knot 41$\pm$3~$\mu$as upstream of the radio core, in agreement with previous detections of upstream emission in BL~Lac \citep{Marscher:2008ii}. The radio core would then correspond to a recollimation shock at $\sim$40~$\mu$as from the jet apex, part of a pattern that also includes two other recollimation shocks at approximately 100 and 250~$\mu$as. Our relativistic magnetohydrodynamic simulations show that such a pattern of recollimation shocks, spaced at progressively larger distance, is expected when the jet propagates through an ambient medium with a decreasing pressure gradient.
  
  Polarization is detected in two components within the innermost 0.5~mas of the core, and in some knots $\sim$3 mas downstream. We have combined the \emph{RadioAstron} 22~GHz image with ground-based observations at 43 and 15~GHz to compute a rotation measure map. Analysis of the core area shows a gradient in rotation measure and Faraday corrected EVPA that depends on the position angle with respect to the core, in similar way as found in the jet of 3C~454.3 \citep{Zamaninasab:2013bn}, and in agreement with numerical RMHD simulations of jets threaded by a helical magnetic field \citep{Broderick:2010dy,Porth:2011ev}.
  
  The stationary feature $\sim$250~$\mu$as from the core contains an enhanced rotation measure and marks a temporary transition in polarization vectors, in which the magnetic field becomes predominantly aligned with the local direction of the jet, in agreement with our RMHD simulations.
  
  We obtain a lower limit for the observed brightness temperature of the unresolved core in our \emph{RadioAstron} image of $2\!\times\!10^{13}$\,K. Using previous estimates of the Doppler factor of $\delta=7.2$ we calculate an intrinsic brightness temperature in excess of $3\times10^{12}$~K, implying that the jet in BL~Lac is not in equipartition of energy between the magnetic field and emitting particles, and suggesting also that its Doppler factor is significantly underestimated.
  
  \emph{RadioAstron} polarimetric space VLBI observations provide a unique tool to study the innermost regions of AGN jets and their magnetic fields with unprecedented high angular resolutions. This will be investigated in a series of papers, results from our \emph{RadioAstron} Key Science Program, which studies a sample of powerful, highly polarized, and $\gamma$-ray emitting blazars.

%% Acknowledgements
\acknowledgements We thank Jos\'e M. Mart\'{\i}, Timothy V. Cawthorne, Oliver Porth, and Carolina Casadio for valuable comments that improved our manuscript. This research has been supported by the Spanish Ministry of Economy and Competitiveness grant AYA2013-40825-P, by the Russian Foundation for Basic Research (projects 13-02-12103, 14-02-31789, and 15-02-00949), and St. Petersburg University research grant 6.38.335.2015. The research at Boston University (BU) was funded in part by NASA Fermi Guest Investigator grant NNX14AQ58G. Y. M. acknowledges support from the ERC Synergy Grant ``BlackHoleCam -- Imaging the Event Horizon of Black Holes'' (Grant 610058). Part of this work was supported by the COST Action MP1104 ``Polarization as a tool to study the Solar System and beyond". The RadioAstron project is led by the Astro Space Center of the Lebedev Physical Institute of the Russian Academy of Sciences and the Lavochkin Scientific and Production Association under a contract with the Russian Federal Space Agency, in collaboration with partner organizations in Russia and other countries. This research is based on observations correlated at the Bonn Correlator, jointly operated by the Max-Planck-Institut f\"ur Radioastronomie (MPIfR), and the Federal Agency for Cartography and Geodesy (BKG). The European VLBI Network is a joint facility of of independent European, African, Asian, and North American radio astronomy institutes. Scientific results from data presented in this publication are derived from the EVN project code GA030B. This research is partly based on observations with the 100 m telescope of the MPIfR at Effelsberg. The VLBA is an instrument of the National Radio Astronomy Observatory, a facility of the National Science Foundation operated under cooperative agreement by Associated Universities. The relativistic magnetohydrodynamic simulations were performed on Pleiades at NASA and on LOEWE at the Goethe University Frankfurt.

{\it Facilities:} \facility{\emph{RadioAstron} Space Radio Telescope (Spektr-R), EVN, VLBA}

%% Bibliography
%\begin{thebibliography}{}
\bibliography{Referencias}{}

%\end{thebibliography}

\begin{table*}
\caption{Gaussian model fits for the VLBA-BU-BLAZAR 43~GHz data}
\label{Tb:43_mf}
\begin{center}
\begin{tabular}{c|c|c|c|c}
  \hline
  Epoch  &      Flux       &     Distance    &        PA       &      Size       \\
  (year) &      (Jy)       &      (mas)      &   ($^{\circ}$)  &     (mas)       \\
 \hline
 2013.96 & 1.620$\pm$0.087 &     \ldots      &    \ldots       & 0.044$\pm$0.005 \\    
         & 2.100$\pm$0.111 & 0.096$\pm$0.005 &   $-174\pm$5\pz & 0.062$\pm$0.006 \\    
         & 0.620$\pm$0.039 & 0.285$\pm$0.005 & \pp$178\pm$1\pz & 0.087$\pm$0.007 \\    
         & 0.227$\pm$0.021 & 0.425$\pm$0.014 & \pp$179\pm$2\pz & 0.132$\pm$0.005 \\    
         & 0.049$\pm$0.010 & 1.260$\pm$0.123 &   $-167\pm$6\pz & 0.370$\pm$0.019 \\    
         & 0.084$\pm$0.016 & 1.660$\pm$0.057 &   $-170\pm$2\pz & 0.256$\pm$0.012 \\    
         & 0.129$\pm$0.020 & 2.590$\pm$0.100 &   $-173\pm$2\pz & 0.507$\pm$0.025 \\    
         & 0.097$\pm$0.020 & 3.790$\pm$0.300 &   $-166\pm$6\pz & 1.390$\pm$0.070 \\    
 \hline
 2014.05 & 0.547$\pm$0.028 &      \ldots     &    \ldots       &        $<$0.005 \\        
         & 1.040$\pm$0.059 & 0.124$\pm$0.005 &   $-170\pm$2\pz & 0.063$\pm$0.005 \\
         & 1.110$\pm$0.063 & 0.304$\pm$0.005 &   $-179\pm$1\pz & 0.104$\pm$0.005 \\
         & 0.168$\pm$0.018 & 0.443$\pm$0.018 & \pp$178\pm$2\pz & 0.140$\pm$0.007 \\
         & 0.097$\pm$0.017 & 1.530$\pm$0.078 &   $-169\pm$3\pz & 0.356$\pm$0.018 \\
         & 0.054$\pm$0.011 & 1.960$\pm$0.100 &   $-168\pm$3\pz & 0.322$\pm$0.016 \\
         & 0.043$\pm$0.009 & 2.480$\pm$0.084 &   $-176\pm$2\pz & 0.255$\pm$0.012 \\
         & 0.105$\pm$0.019 & 2.990$\pm$0.150 &   $-172\pm$3\pz & 0.652$\pm$0.033 \\
 \hline
 2014.15 & 0.785$\pm$0.045 &      \ldots     &    \ldots       & 0.028$\pm$0.005 \\     
         & 0.735$\pm$0.043 & 0.113$\pm$0.005 &   $-166\pm$3\pz & 0.048$\pm$0.005 \\
         & 0.434$\pm$0.030 & 0.310$\pm$0.006 &   $-176\pm$1\pz & 0.091$\pm$0.005 \\
         & 0.070$\pm$0.015 & 0.535$\pm$0.054 & \pp$172\pm$6\pz & 0.224$\pm$0.011 \\
         & 0.010$\pm$0.003 & 1.050$\pm$0.088 & \pp$172\pm$5\pz & 0.126$\pm$0.007 \\
         & 0.021$\pm$0.005 & 1.390$\pm$0.087 &   $-179\pm$4\pz & 0.182$\pm$0.010 \\
         & 0.058$\pm$0.015 & 1.580$\pm$0.092 &   $-170\pm$3\pz & 0.318$\pm$0.016 \\
         & 0.065$\pm$0.016 & 1.960$\pm$0.100 &   $-170\pm$3\pz & 0.357$\pm$0.018 \\
         & 0.037$\pm$0.010 & 2.520$\pm$0.130 &   $-175\pm$3\pz & 0.332$\pm$0.016 \\
         & 0.133$\pm$0.020 & 3.180$\pm$0.160 &   $-172\pm$3\pz & 0.755$\pm$0.038 \\
 \hline 
 2014.34 & 1.450$\pm$0.078 &      \ldots     &     \ldots      & 0.026$\pm$0.005 \\
         & 1.510$\pm$0.082 & 0.140$\pm$0.005 &   $-161\pm$2\pz & 0.049$\pm$0.006 \\
         & 1.290$\pm$0.071 & 0.297$\pm$0.005 &   $-162\pm$1\pz & 0.054$\pm$0.007 \\
         & 0.739$\pm$0.045 & 0.335$\pm$0.006 &   $-172\pm$1\pz & 0.113$\pm$0.006 \\
         & 0.027$\pm$0.007 & 0.658$\pm$0.022 & \pp$176\pm$1\pz & 0.031$\pm$0.005 \\
         & 0.013$\pm$0.003 & 1.120$\pm$0.113 & \pp$171\pm$6\pz & 0.175$\pm$0.008 \\
         & 0.030$\pm$0.008 & 1.310$\pm$0.071 & \pp$178\pm$3\pz & 0.184$\pm$0.010 \\
         & 0.111$\pm$0.018 & 1.640$\pm$0.077 &   $-176\pm$3\pz & 0.374$\pm$0.020 \\
         & 0.025$\pm$0.006 & 2.290$\pm$0.100 &   $-167\pm$3\pz & 0.224$\pm$0.011 \\
         & 0.040$\pm$0.010 & 2.650$\pm$0.110 &   $-173\pm$2\pz & 0.305$\pm$0.015 \\
         & 0.141$\pm$0.021 & 3.670$\pm$0.200 &   $-171\pm$3\pz & 0.967$\pm$0.048 \\
 \hline
 2014.47 & 0.991$\pm$0.053 &     \ldots      &      \ldots     &        $<$0.005 \\       
         & 0.902$\pm$0.052 & 0.148$\pm$0.005 &   $-165\pm$2\pz & 0.058$\pm$0.005 \\
         & 1.180$\pm$0.066 & 0.335$\pm$0.005 &   $-164\pm$1\pz & 0.066$\pm$0.005 \\
         & 0.055$\pm$0.011 & 0.437$\pm$0.015 &   $-156\pm$1\pz & 0.047$\pm$0.006 \\
         & 0.104$\pm$0.017 & 0.634$\pm$0.082 &   $-177\pm$7\pz & 0.383$\pm$0.019 \\
         & 0.030$\pm$0.007 & 0.961$\pm$0.070 & \pp$167\pm$4\pz & 0.176$\pm$0.009 \\
         & 0.049$\pm$0.012 & 1.190$\pm$0.060 & \pp$174\pm$3\pz & 0.202$\pm$0.010 \\
         & 0.084$\pm$0.016 & 1.520$\pm$0.080 &   $-177\pm$3\pz & 0.332$\pm$0.026 \\
         & 0.047$\pm$0.012 & 1.980$\pm$0.200 &   $-175\pm$6\pz & 0.524$\pm$0.023 \\
         & 0.044$\pm$0.011 & 2.660$\pm$0.170 &   $-173\pm$4\pz & 0.459$\pm$0.023 \\
         & 0.128$\pm$0.021 & 3.740$\pm$0.220 &   $-171\pm$3\pz & 0.960$\pm$0.048 \\
 \hline
\end{tabular}
\end{center} {\bf Notes.} Tabulated data correspond to flux density, distance and position angle from the core, and size. Errors in the model-fit parameters for each component are estimated based on its brightness temperature following \cite{2015ApJ...813...51C}.
\end{table*}

\end{document}